\documentstyle[preprint,aps]{revtex}

\def\SSC#1{{\scriptstyle {\sc
#1}}}

\begin{document}
\draft

\title{Pressure dependence of the sound velocity in a 2D lattice of
Hertz-Mindlin balls: a mean field description}
\author{B. Velick\'y\cite{Charles} and C.  Caroli}
\address{Groupe de Physique des Solides\cite{CNRS}, 2 place Jussieu,
75251 Paris Cedex 05, France.}

\date{\today}

\maketitle

\begin{abstract}
We study the dependence on the external pressure $P$ of the velocities
$v_{L,T}$ of long wavelength sound waves in a confined 2D h.c.p.
lattice of 3D elastic frictional balls interacting via one-sided
Hertz-Mindlin contact forces, whose diameters exhibit mild
dispersion. The presence of an underlying long range order enables us
to build an effective medium description which incorporates the
radial fluctuations of the contact forces acting on a single site. Due
to the non linearity of Hertz elasticity, self-consistency results in a
highly non-linear differential equation for the "equation of state"
linking the effective stiffness of the array with the applied
pressure, from which sound velocities are then obtained. The results
are in excellent agreement with existing experimental results and
simulations in the high and intermediate pressure regimes. It emerges
from the analysis that the departure of $v_{L}(P)$ from the ideal $P^{1/6}$
Hertz behavior must be attributed primarily to the fluctuations of
the stress field, rather than to the pressure dependence of the number of
contacts.
\end{abstract}


\section{Introduction}
Sound propagation in a dry confined granular medium still is, to a
large extent, an open question. In particular, a long-standing puzzle
is concerned with explaining the dependence of sound velocities on the
externally applied pressure $P$.

The load bearing intergrain contacts which ensure the mechanical
stability of the packing are of the Hertz type, i.e. their
longitudinal stiffness (along the intercenter axis) scales as
$F^{1/3}$, with $F$ the corresponding load.  As shown by Mindlin\cite{KLJ},
this scaling also holds, for frictional balls, for their shear stiffness,
provided that the shear load borne by the contact is much smaller than the
friction threshold. So, one intuitively expects that the velocity of, say,
longitudinal sound $v_{L}\sim P^{1/6}$.

However, experimental results depart strongly from
this expectation : $v_{L}$ is found to exhibit a much faster
$P$-dependence, which is commonly characterized by
"effective exponents" $\nu = {\rm d}(\log v)/{\rm d}(\log P)$. Values of
$\nu$ of order roughly $1/4$ are often mentioned. Such a behavior is
observed for 3D random grain packings, which present what we will term
"strong topological disorder" but also, more surprisingly, for artificially
built regular arrays of quasi-monodisperse balls. This was first shown on a
3D f.c.c. lattice by Duffy and Mindlin\cite{Duffy}, who found that the
faster dependence with $\nu \sim 1/4$ in the intermediate pressure
range tended asymptotically, at high $P$'s, towards the Hertz
$P^{1/6}$ dependence. Recently, Coste and Gilles\cite{Gilles} have
investigated in detail sound propagation in a 2D hexagonal lattice of steel
balls. Their experimental results for $v_{L}(P)$, which are qualitatively
similar to those of Duffy and Mindlin, have motivated the present study.

Various explanations for this behavior for strongly disordered packings
have been put forward. In particular, Goddard\cite{Goddard}
proposed that a $1/4$ value of $\nu$ might originate from the existence of
conical contacts, while de Gennes\cite{de Gennes} suggested that the
presence of heterogeneous shells surrounding the grain bodies might in some
cases be relevant. A possibly more natural explanation lies in the pressure
dependence of the numberof load bearing contacts in the packing. It has
been at the center of several recent works. In particular, a direct
numerical study by Makse {\em et al.}\cite{Makse} demonstrates the
correlation between number of contacts and sound velocity in a 3D system.

For regular arrays, such an explanation may at first appear
doubtful. However, J.N. Roux\cite{Roux1} has studied numerically
the 2D h.c.p.  structure and found that even a minute dispersion
in ball radii leads to a similar effect. Namely, due to purely
geometrical constraints, as the pressure increases, the average number of
contacts per ball $N_{c}$ varies from $\sim 2.5$ at the rigidity threshold
to its maximum value of $6$ at high $P$.

Although such effects seem to be within the reach of a mean field
description, the various attempts in this direction have up to now
failed to account for deviations from the Hertz power law
\cite{Digby}, \cite{Walton}, \cite{Winkler}.
As already suggested by Makse {\em et al.}\cite{Makse}, we believe
this to result from the
central but implicit assumption of these models that the local contact
arrangement deforms homothetically when $P$ changes.  This amounts to
neglecting the essential effect of local stress inhomogeneities: due to the
elastic Hertz deformations, they result in a dispersion of intercenter
distances, and hence of the bond strengths. In other words, even in the
absence of a change in contact number, the change of bond stiffness induced
by a pressure change is non affine.

While it is certainly very difficult to improve upon this approximation in
the case of strong topological disorder, the case of a periodic array
of weakly disperse balls seems amenable to a realistic
description. Indeed, in this case, the existence of a reference
lattice permits to formulate a mean field theory in the spirit
of the single center self-consistent CPA approach\cite{Sheng},
developed already long ago
to describe the electronic properties of metallic alloys.
Such a route was already explored in a series of articles on
depleted elastic networks\cite{FengMF1}, \cite{FengMF2}.
However, in these works, the distribution of active contacts was
assumed to be known and independent of the external stress, and
the self-consistent condition was formulated in terms of a single bond
approximation, in the spirit of Kirkpatrick's approach to percolation
problems\cite{Kirkpatrick}.

In this paper, we build an effective medium description of a 2D h.c.p.
array of Hertz-Mindlin balls which does account for local deformation
due to the disorder in ball radii. In contrast with previous theories,
our self-consistency condition does depend on the global external
stress.  Clearly, in such a system, the higher the external pressure,
 the smaller the relative disorder. So, our mean field appears as a high-$P$
expansion.
It therefore complements the numerical studies of Roux, which deal
with the low-$P$ regime where percolation effects are dominant.

We show that our predictions account quantitatively for the experimental
results of Gilles and Coste. Moreover, comparison with Roux's results
in an overlapping intermediate pressure range allows us to
determine the range of validity of our effective medium approach. We
find that it holds down to pressures where $N_{c}$ has decreased by
about $15\% $ of its saturation value. From all this, we conclude
that the basic physical effect responsible for departures from the
$P^{1/6}$ law is, rather than the variation of contact number in
itself, the  disorder induced spatial stress fluctuations.

The article is organized as follows. In Section II we set
the basis of our model by writing the dynamical equations for a set
of Hertz-Mindlin contacts under equilibrium forces aligned with
the intercenter directions, and solve them for an ideal lattice
of perfectly identical balls. In Section III, we build up our mean field
description, apply it to the h.c.p. lattice, and obtain from it
the equation of state, i.e. the force-displacement relation from which
the effective bulk and shear moduli are derived. Section IV compares
in detail the mean field predictions with the experimental and
numerical results.

\section{Basic model and dynamical
equations}
\subsection{Equations of motion}
Let us first consider two
spherical balls, labeled $(i), (j)$, of radii $d_{i}, d_{j}$, made of the
same material of Young modulus $E$, Poisson ratio $\sigma$, density $\rho$.
The balls at equilibrium are in Hertz contact\cite{KLJ} under a force $f$
directed along the intercenter axis $ij$, i.e. normal to the contact circle.

The normal force $f$ displaces the intercenter distance from
$d_{ij} = {\frac{1}{2}}(d_{i}+d_{j})$ to $a_{ij} = d_{ij}-\delta_{ij}$, and:
\begin{equation}
\label{eq:1}
\delta_{ij}(f)
= (\frac {9 f^{2}}{4d^{*}_{ij}E^{*2}})^{1/3}
\end{equation}

Due to solid friction, when submitted to a tangential force smaller
than the friction threshold $ \mu _{s}f$, the contact is pinned and
cannot slide. The elastic response of this system to small additional
forces $\delta f_{x}, \delta f_{y}$ in the $(x, y)$ plane is
described, in the linear approximation, by two stiffness
coefficients, given by the Hertz-Mindlin expressions, namely \cite{KLJ}:
\begin{enumerate}
\item compression (normal)
stiffness
\begin{equation}
\label{eq:2}
k_{ij} = - \frac{\delta
f_{x}}{\delta x_{ij}} =
({\frac{3}{2}}E^{*2}d_{ij}^{*}
f)^{1/3}
\end{equation}
where
\begin{equation}
\label{eq:3}
E^{*} = \frac{E}{2(1-\sigma^{2})}
\,\,\,\,\,\,\,\,\,\,\,
(d_{ij}^{*})^{-1} =
{\frac{1}{2}}(d_{i}^{-1} +
d_{j}^{-1})
\end{equation}
\item shear (tangential) stiffness :
\begin
{equation}
\label{eq:4}
\kappa_{ij} = \eta k_{ij} \,\,\,\,\,\,\,\,\,\,\,\,
\eta =
\frac{2(1-\sigma)}{2-\sigma}
\end{equation}
\end{enumerate}

Let us insist here that the fact that the ratio $ \eta = \kappa_{ij}/k_{ij}$ is
a mere material parameter, independent of the values of the equilibrium
forces, only holds for the case of a purely normal equilibrium loading,
which we assume here to be the case.
In the general situation with a finite equilibrium tangential load $f_{t}$,
$\eta \rightarrow \eta[1 - (f_{t}/\mu_{s}f)^{1/3}]$. In all the
following, we will restrict ourselves to situations where all
(equilibrium and non-equilibrium) interball forces
are near normal, so that both the static and dynamic shear responses are
described by the linear shear stiffness (equation (\ref{eq:4})).

Small displacements of the balls can be decomposed into rigid
translations $\bf{u}_{i}, \bf{u}_{j}$, and rotations about $Oz$ by angles
$\phi_{i},\phi_{j}$. Denoting $\hat{\bf n}_{ij},\hat{ \bf t}_{ij}$ the
unit vectors normal and tangent to the $(ij)$ contact with $\hat{\bf
n}_{ij}$ directed  from $(i)$ to $(j)$  (see Figure \ref{fig:1}), one
immediately finds that the force and torque on ball $(i)$ associated
with the $(ij)$ contact read respectively:
\begin{equation}
\label{eq:5}
\begin{array}{rcl}
\delta{\bf F}_{i,j} &=&{\ }{\ } 
k_{ij}[({\bf u}_{i}-{\bf u}_{j}).\hat{{\bf n}}_{ij}]\hat{\bf n}_{ij}\\
&&+ \eta  k_{ij}[({\bf u}_{i}-{\bf u}_{j}).
\hat{{\bf t}}_{ij}]\hat{{\bf t}}_{ij}\\
 & &-\eta
 k_{ij}\frac{1}{2} (d_{j}\phi_{j}+d_{i}\phi_{i})\hat{\bf t}_{ij}
 \end{array}
 \end{equation}  
\begin{equation}
\label{eq:6}
\delta
C_{i,j} = \eta k_{ij}\frac{d_{i}}{2}  [({\bf u}_{j} - {\bf
u}_{i}).
\hat{{\bf t}}_{ij} - \frac{1}{2}
(d_{j}\phi_{j}+d_{i}\phi_{i})]
\end{equation}
and, in the small displacement limit appropriate to sound
propagation, the equations of motion for a 2D lattice of balls are:
\begin{equation}
\label{eq:7}
M_{i}\ddot{{\bf u}}_{i} = \sum
\limits_{\{j\}}
\delta{\bf F}_{i,j}\,\,\,\,\,\,\,\,\,\,\,\,\,\, M_{i} =
\frac{\pi\rho
d_{i}^{3}}{6}
\end{equation}
\begin{equation}
\label{eq:8}
I_{i}\ddot {\phi}_{i} = \sum \limits_{\{j\}}
\delta C_{i,j}\,\,\,\,\,\,\,\,\,\,\,\,\,\, 
I_{i} = \frac{ d_{i}^{2} M_{i}}{10}
\end{equation}
where the sums are restricted to nearest neighbors
in direct contact with $(i)$.

\subsection{Ideal h.c.p.
lattice}
We now consider the ideal case of a 2D h.c.p. lattice of balls of
equal diameters $d$ prepared so that, at equilibrium, the interball
forces, of magnitude $f$, are directed along the normals to the contacts
$\hat{ \bf{n}}_{i}$ (see Figure \ref{fig:1}).
Such a "hydrostatic" configuration
can be realized by applying a force per unit length $P = f{\sqrt 3}/d $ on
a hexagonal container with walls along the dense ball rows, as
realized in \cite{Gilles}. The unit cell is defined by the two
vectors $ {\bf a}_{1,2} = d \hat{\bf n}_{1,2}$, the ball centers
by ${\bf R}_{mn} = m{\bf a}_{1}+ n {\bf a}_{2}$.

One then obtains from equations (\ref{eq:5}) - (\ref{eq:8}), for the
vibrational modes of this system:
\begin{equation}
\label{eq:9}
{\bf u}_{mn} = {\bf u} e^{i({\bf
q}.{\bf R}_{mn}-\omega t)} \,\,\,\,\,\,\,\,
\phi_{mn} = \phi e^{i({\bf q}.{\bf R}_{mn}-\omega t)}
\end{equation}
\begin{equation}
\label{eq:10}
\begin{array}{r@{}l@{}l}
M\omega^2{\bf u} =&&\\
2k&\sum\limits_{p=1}^{3}&(1 - C_{p}({\bf q}) 
[({\bf u}.\hat{\bf n}_{p})\hat{\bf n}_{p} + \eta {\bf
u}.\hat{\bf t}_{p})\hat{\bf t}_{p}]\\ 
+ i\eta kd\,\phi&\sum\limits_{p=1}^{3}&S_{p}({\bf q})\,\hat{\bf t}_{p}
\end{array}
\end{equation}
\begin{equation}
\label{eq:11}
I\omega^2{\phi} = -
i \eta k d \sum\limits_{p=1}^{3}S_{p}({\bf q}) ({\bf
u}.\hat{\bf t}_{p})
\,+\, \eta k \frac{d^{2}}{2}
\phi\sum\limits_{p=1}^{3}(1+C_{p}({\bf q}))
\end{equation}
where $k$ is the normal stiffness common to all contacts, and:
\begin{equation}
\label{eq:12}
S_{p}({\bf q}) = \sin[d({\bf q}.\hat{\bf n}_{p})]\,\,\,\,\,\,\,\,
C_{p}({\bf q}) = \cos[d({\bf q}.\hat{\bf n}_{p})]
\end{equation}
and use has been made of relations such as $\sum\limits_{p=1}^{6}
\hat{\bf n}_{p} = 0$.

The exact spectrum in the full Brillouin zone, computed for ${\bf q}$ along
two directions of high symmetry, is shown on Figure \ref{fig:2}. It can be
inferred from these results that, in this close-packed lattice, the
anisotropy of the spectrum is small.

In the elastic continuum, long wavelength limit $qd \ll 1$, decomposing
the translation amplitude into its longitudinal and transverse
components : ${\bf u} = u_{\parallel}\hat{\bf q} +
u_{\perp}(\hat{\bf z}\wedge\hat{\bf q})$, one finds that the vibration
spectrum, which is isotropic due to the hexagonal symmetry of the lattice,
is composed of three branches:
\begin{enumerate}
\item pure longitudinal acoustic modes of frequency $\omega_{T} = v_{L}q$,
where the longitudinal sound velocity $v_{L}$ is given by:
\begin{equation}
\label{eq:13}
v_{L}^{2} =
\frac{9}{8}(1+\frac{\eta}{3})\frac{k}{M}d^{2}
\end{equation}
\item Two branches of mixed modes containing both a transverse
translational and a rotational component. One of them is acoustic :
$\omega = v_{T}q$, where the transverse sound velocity reads:
\begin{equation}
\label{eq:14}
v_{T}^{2} = \frac{3}{8}
(1+\eta)\frac{k}{M} d^{2}
\end{equation}
The second, which corresponds to pure rotation in
the $q = 0$ limit, is an optical branch, defined by:
\begin{equation}
\label{eq:15}
\omega_{R}^{2} = 30\eta
\frac{k}{M}(1-\frac{1}{20}q^{2}d^{2})
\end{equation}
\end{enumerate}

That is, as already shown by Schwartz {\em et al.} \cite{Fenghcp}, the
specificity of the vibration spectra of granular systems in frictional
Hertz contact, as compared with atomic systems, lies in
the additional degree of freedom associated with ball rotation. This
remains coupled, in the long wavelength limit, with shear deformation,
leading to a contribution $(- 3\eta kd^{2}/4M)$ to $v_{T}^{2}$.

Also shown on Figure \ref{fig:2} are the long wavelength dispersion curves.
They can be seeen to provide a good approximation for the exact
spectrum in a sizeable fraction of the Brillouin zone.

It is worth recalling, at this point, that our equations
of motion cease to be valid when the frequency approaches that of the
lowest acoustic resonance of a
ball : $\omega_\SSC{res} \approx v_\SSC{bulk}/d$, with $v_\SSC{bulk}$ a sound
velocity of the material constituting the balls. Indeed, in this
situation, elastic deformations are no longer localized in a
small region of extension on the order of the contact radius, internal
deformations become important, and the restoring forces can no longer
be described simply via the Hertz-Mindlin stiffnesses.  Roughly
speaking, this means that our expressions for the acoustic branches
of the spectrum are valid provided that $v_{L,T} \ll
v_\SSC{bulk}$. In view of equations (\ref{eq:13}), (\ref{eq:14}) and
(\ref{eq:2}),  this simply amounts to $(f/Ed^{2})^{1/6} \ll 1$, which is
realized under ordinary experimental conditions \cite{plastic}. Note that
this condition is equivalent to stating that  the radius of the Hertz
contact circle must be much smaller  that the ball radius, which is
precisely the condition for the Hertz approach to hold.

From planar continuum elasticity applied to a medium with hexagonal 
symmetry, the velocities of sound waves with propagation and
polarization directions in the basal plane read :
\begin {equation}
\label{eq:16}
v_{L}^{2} =
\frac{K +G}{\tilde\rho}\,\,\,\,\,\,\,\,\,\,\, \,
v_{T}^{2} =
\frac{G}{\tilde\rho}
\end{equation}
with $\tilde\rho$ the mass density of the medium, $K$ and $G$ its bulk
and shear moduli.

Comparison between these expressions and
equations (\ref{eq:13}), (\ref{eq:14}) enables us to define elastic
moduli for our ball lattice - a result which will be of use
in the disordered case. They read :
\begin{equation}
\label{eq:17}
K =
\frac{\sqrt{3}}{2}\frac{k}{d}\,\,\,\,\,\,\,\,\,\,\,\,\,G =
\frac{(1 +
\eta)}{2}\,K
\end{equation}
where $\tilde\rho$ is related to the density of the ball material by:
 $\tilde\rho = \pi \rho /3\sqrt{3}$.

\section{Disordered lattice}

\subsection{Random h.c.p.
array of balls under hydrostatic compression}
Consider an ensemble of balls whose
material parameters are identical, while their diameters vary at
random  with a continuous or discrete statistical distribution.
A diameter value $d^Q$, with a formal label $Q$,
has the probability, or fractional concentration, $c^Q$, and $\sum c^Q=1$.

The mean diameter $d$ of a set of $N\gg 1$ such balls may be obtained by
 configuration average, which we will denote by $\langle\ldots\rangle$:
\begin{equation}
\label{eq:18}
d={1 \over
N}\sum\limits_i d_i\longrightarrow d\equiv\langle d\rangle=
\sum\limits_Q
c^Q\,d^Q
\end{equation}
We  define the random deviations by
\begin{equation}
\label{eq:19}
\begin{array}{rclcrcl}
{\mit \Delta}_i&=&d_i-d,&{\mit \Delta}^Q&=&d^Q-d\\
\langle{\mit \Delta}\rangle&=&0,&{\mit \Delta}_\SSC{rms}^2\equiv
\langle{\mit \Delta}^2\rangle&=&\sum
c^Q({\mit \Delta}^Q)^{2}
\end{array}
\end{equation}
In actual computations, we will use the uniform distribution with
full width $W=\max \{{\mit \Delta}^Q\} -\min
\{{\mit \Delta}^Q\}$, which means
${\mit \Delta}_\SSC{rms}=W\sqrt{12}$.

One sample (configuration) of the random h.c.p. array of balls may be created
by randomly distributing the balls in an uncorrelated way over the sites of a
reference h.c.p.  lattice. The number of balls is assumed to be sufficiently
large for the thermodynamic limit to be approached.
The sample is then random, but macroscopically homogeneous.
The balls are brought into contact and further compressed by external
compressive forces applied to the sample boundaries so that the average internal
stress is hydrostatic. An easy way of achieving this is to
assume a hexagon shaped sample and to apply to all its sides
the same macroscopic pressure force. The pressure force per unit length $P$
will be denoted henceforth {\em linear pressure} .  Under this
pressure, the size of the compressed sample is reduced, while the symmetry
of the lattice is preserved on average.  It is thus meaningful to introduce
an {\em average lattice spacing $a$} as a macroscopic parameter having a
thermodynamic limit and globally characterizing the state of the sample.

The pressure and the size of the compressed system are related by
{\em the macroscopic equation of state} for the random lattice under
hydrostatic compression.  It will be convenient to introduce it in the form

\begin{equation}
 \label{eq:20}
 f=\tilde{f}(a),\quad f=P\cdot d/\sqrt(3)
\end{equation}
 where $\tilde{f}$ is the functional dependence in question and $f$
denotes the {\em average intergrain hydrostatic force} associated with
the linear pressure $P$.  The {\em effective normal stiffness $\tilde{k}$}, 
and the effective bulk modulus $K_\SSC{eff}$ are then given by
\begin{equation}
\label{eq:21}
\tilde{k}(a)=-{{{\rm d}\tilde{f}(a)} \over {{\rm d}a}},\quad
K_\SSC{eff}\cdot d=-{{a^2} \over {\displaystyle \left({{{\rm d}a^2}
\over {{\rm d}P}}\right)}}={{\sqrt{3}} \over 2}\cdot\tilde{k}
\end{equation}
As for the $d$ factor on the left hand side, {\em cf.} Eq. (\ref{eq:17}).

While the global characteristics,  $P$ and $a$, have
thermodynamic limits, the ball positions and contact forces are
subject to pronounced fluctuations, in particular for small external
loads.  In experiments, this double nature of the disordered state
is manifested by the coexistence of coherent signals and irregular speckles
in the acoustic response \cite{Gilles}, \cite{Jia}, in numerical
simulations most clearly by the formation of force chains \cite{Roux1},
\cite{Radjai}.

Let us first verify that the geometrical disorder can be taken as small,
as demanded is Sec. IIA.  If the relative diameter spread is small,
the equilibrium disordered system can be treated, for any
external pressure, as a distorted lattice.  The balls are slightly
displaced, and, in the case of frictional balls, the contact points may be
somewhat off the intercenter lines, which gives rise to non-normal contact
forces. Taking as representative the data of \cite{Gilles},
$W \lesssim\ 8\,\mu$m, $d=8\,$mm for steel balls,
we see that the relative dispersion in "bond" lengths is of the order of
$10^{-3}$. By geometrical considerations, this corresponds to deviations of
the direction of the contact force from
the normal also of about  $10^{-3}$ radian. These figures indicate that the
geometrical disorder in this case is small indeed.

Fluctuations in the magnitude of the random contact forces, by contrast,
may be quite large.  By Eq.  (\ref{eq:1}), the contact force as a function of
intercenter distance is given by
\begin{equation}
\label{eq:22}
f_{ij}(x)=\left\{\begin{array}{ll}
{\textstyle {2 \over 3}}E^*(\underbrace{d_{ij}^*}_
{\textstyle \approx d})^{1 \over 2}
(d_{ij}\,-\,x)^{3 \over 2} &\qquad d_{ij}\,-\,x>0\\
0&\qquad
d_{ij}\,-\,x<0
\end{array}\right.
\end{equation}
As indicated by the underbrace in (\ref{eq:22}),
it is consistent to neglect the randomness of $d_{ij}^*$.
Namely, the fractional fluctuation  of
$d_{ij}^*\approx d +{1 \over 2}{{\mit \Delta}_i+{\mit \Delta}_j}$
is $\sim {\mit \Delta}_\SSC{rms}/d$.

On the other hand, the disorder of  the last factor in
(\ref{eq:22}) is crucial: for the Hertz displacement
$d_{ij}\,-\,x$, the  ratio  ${\mit \Delta}_\SSC{rms}/(d-x)$
may be large or small depending on the degree of compression of the balls,
while
both basic conditions, ${\mit \Delta}_\SSC{rms}\,\ll\,d$  (small disorder),
and $d\,-\,x\,\ll\,d$ (Hertzian picture), remain satisfied.
We may thus envisage three different regimes:

\bigskip
\begin{center}
\begin{tabular}{|c|c|}
\hline
regime&condi
tion\\ \hline
low pressure&${\mit \Delta}_\SSC{rms}\,\gg\,d-x$\\
intermediate
pressure&${\mit \Delta}_\SSC{rms}\,\approx\,d-x$\\
high
pressure&${\mit \Delta}_\SSC{rms}\,\ll\,d-x$ \\
\hline
\end{tabular}
\end{center}
\bigskip

In the high pressure Hertzian regime, ${\mit \Delta}_\SSC{rms}\,\ll\,
d\,-\,x\,\ll\,d $, the disorder appears as a perturbation of a basic
homogeneous and homogeneously compressed crystal.  This natural conjecture
has given rise to two simple, but useful approximations (see \cite{Roux1})
for the true effective medium:
\begin{enumerate}
\item The {\em Averaged Lattice Approximation}
(ALA) replaces the true sample by an ideal lattice of balls with the
diameter $d$. The constitutive law and the bulk modulus thus read\cite{plus}
\begin{equation}
\label{eq:23}
\tilde{f}(a)={\textstyle {2 \over 3}}E^*d ^{1 \over 2}
(d\,-\,a)^{3 \over 2}_{_+}\equiv F(a)
\end{equation}
\begin{equation}
\label{eq:24}
K_\SSC{ala}=E^*\left({9 \over {16}} {P \over {E^*d}} \right)^{1 \over 3}
\end{equation}
Thus, the bulk modulus naturally obeys the plain Hertz ${1 \over 3}$ law,
as is appropriate for a periodic array (see Sec. II.).
\item The {\em  Averaged Force Approximation} (AFA) assumes the
balls to be positioned at any pressure exactly at
the lattice sites, while the contact forces are given by (\ref{eq:22}). In
this strictly symmetric geometry, no shear forces between the grains occur,
and the effective hydrostatic force between two adjacent sites
is obtained by  configuration averaging,
\begin{equation}
\label{eq:25}
\begin{array}{rcl}
\tilde{f}(a)&=&\langle f_{ij}(a)\rangle\equiv F_\SSC{ave}(a)\\
F_\SSC{ave}(a)&=&\sum\limits_{Q,Q'}\,c^Qc^{Q'}F^{QQ'}(a)
\end{array}
\end{equation}
where we have introduced the notation:
$$
\label{eq:25a}
\begin{array}{rcl}
F^{QQ'}(x)&=&{\textstyle {2 \over
3}}E^*d ^{1 \over 2}(d^{QQ'}\,-\,x)^{3 \over 2}_{_+}
\end{array}
\eqno(25a)
$$
for a force at a separation $x$ between two balls of the
prescribed species $Q,\ Q'$ with $d^{QQ'}={1 \over 2}(d^{Q}+d^{Q'}) = d
+{\mit \Delta}^{QQ'}$.

Asymptotically,
\begin{equation}
\label{eq:26}
\tilde{f}(a)\sim
F(a)\cdot\left(\,1\,+\,{3 \over 8}{{\mit \Delta}_\SSC{rms}^2
\over {(d\,-\,a)^2}}\right)
\end{equation}
and the bulk modulus becomes
\begin{equation}
\label{eq:27}
K_\SSC{afa}=K_\SSC{ala}\cdot\left(\,1\,-{{\,1\,} \over 8}\,  
{{{\mit \Delta}_\SSC{rms}^2\cdot
({4 \over
3} {E^*}^2d^{-1})^{{2 \over 3}}} \over {P^{4 \over
3}}}\right)
\end{equation}
These two equations show that the AFA effective contacts are
non-Hertzian, and that the  pressure dependence of  the elastic modulus
deviates from the simple ${1 \over 3}$ power law.
\end{enumerate}
While the ALA has an arbitrary nature, the AFA seems to be
justified for high pressures, when the geometrical disorder is small
compared to the compressive displacements, while the contact forces still
continue to exhibit random fluctuations. These force fluctuations then
appear as being responsible for the non-hertzian features of the effective
contacts\cite{approx}.

At lower, "intermediate", pressures, according to our above classification,
all grains are still mutually
engaged through their contacts, but the random displacements become
comparable to the global compressive deformations. The AFA is then not
sufficient, while an approximation taking these local lattice distortions
into account in an averaged manner may be satisfactory. The EMA described
in the next section is an attempt in this direction.

\subsection{Effective Medium Approximation}
Now, we are ready to build our Effective Medium Approximation for the
disordered hexagonal array of balls. We use this name, conventional in the
context of granular assemblies, but, as sketched already in the
Introduction, a more descriptive name would refer to the mean-field nature
of the approximation, or, alternatively, to its "single site" character.
The universal idea of such approximations is as follows \cite{Sheng}.
We will only consider disordered systems in which a periodic geometrical
lattice (2D h.c.p. in our case) is filled by elementary objects associated
with individual sites (balls for us) and having randomly variable
characteristics (radii). It is assumed that the macroscopic properties of such
random systems can be obtained, in principle, by configuration averaging.
The configurationally averaged system is exactly periodic again.
In the mean-field approximation, one assumes that it can be represented
as a periodic array of effective elementary objects similar in nature to
the random elements of the real array. The characteristics of
these are determined by the following self-consistent
procedure. One of the effective elements is replaced back ("substituted")
by a true random one. The new system is locally sampled in a random
fashion. It is then required that a configuration average of these
locally distorted systems restores the average behavior.  This condition
determines the characteristics of the effective elements.  Once this has
been done, the whole task of configuration averaging is complete.

An effective medium theory along these lines was developed by Feng 
{\em et al.} in \cite{FengMF1} and \cite{FengMF2} to study depleted 
elastic networks on lattices. The basic element in the theory of 
these authors was a single bond connecting
two lattice sites. The bonds were described by their linear stiffnesses,
whose random distribution was prescribed $a\,priori$. The resulting EMA (a
"single-bond Coherent Potential Approximation") provided a theory for 
the effective linear elasticity of the network,
and for its vibrational spectrum.

While this bond-EMA was a successful theory in its own area, we contend that
it cannot be used for random Hertz lattices for two essential reasons:
\begin{enumerate}
\item\quad The elementary object in a granular system is a ball. The star
of (six for 2D h.c.p.)
contacts surrounding the ball, that is of "bonds" stemming from the center,
is statistically correlated, and the bonds cannot be treated as independent.
\item\quad The linear stiffnesses of the contacts are not known beforehand.
They are indirectly specified by the average external pressure, but their
local fluctuations depend on the equilibrium ball
positions  and cannot be determined independently of the non-linear static
equilibration   of the Hertz array at a given pressure.
\end{enumerate}
These two points are of a different nature and importance.
The second point holds for any granular system, and we believe that it
is precisely this which has been the obstacle against developing a
satisfactory EMA for the acoustic response of granular materials. It may be
said that the grain network should fulfill two contradicting roles at the
same time, namely it should constitute the medium for wave propagation,
and, as well, the random scattering field.  We will
see how this basic problem can be overcome in our rather specifically
constructed case.

\subsubsection{EMA for frictionless balls}

To develop an EMA incorporating these features, we consider first the case
of non-frictional Hertz contacts (formally, $\eta = 0$). The averaged array is
assumed to consist of {\em effective balls}, whose diameter is $a$.
The principal assumption is that the
average hydrostatic force $\tilde{f}(a)$ can be interpreted locally as a
contact force between two effective balls, so that
$\tilde{k}(a)=-{{\rm d} \over {{\rm d}a}} \tilde{f}(a)$
is the stiffness of an effective contact between two such balls.
Now, we select one site, "0", as central and substitute a ball with
diameter $d^Q$ for the effective ball. The difference $d^Q-a$ in diameters
will give rise to an elastic deformation of the effective lattice. In other
words, the substituted ball acts as an elastic inclusion. On average, the
deformation should cancel.  This would define the EMA condition, if only
we were able to determine the force between a real and an effective ball.
This is not possible directly, and we propose to overcome this by
considering a cluster consisting of the central ball and a "corona" of
its nearest neighbors, as sketched in Fig.\ref{fig:3}. These neighbors
have a hybrid nature. In other words, the interface between the
effective medium and the inclusion passes through the corona balls.
All contacts between the corona balls  and  the adjacent
effective balls are taken as effective ones, associated with the force
$\tilde{f}$. This type of approximation is based on the mean field
reasoning: the fluctuations of the remote parts of the lattice should
not have a significant effect on the central site.

Inside the cluster, the corona balls might appear as "true" randomly chosen
balls, so that the contact forces would
be of the form $F^{QQ'}$ as given by Eq.(25a). %
A straightforward procedure involving averaging over the individually
equilibrated positions of such randomly composed clusters of 7
balls, while conceivable, would be disproportionately clumsy.

We prefer to introduce a {\em model of the corona}, which, while simple and
transparent, captures the core features of the problem:
\begin{itemize}
\item
The contact forces between the central ball and the corona balls are
averaged over the corona configurations:
\begin{equation}
\label{eq:28}
f_{0i}(x)\rightarrow
F^{Q}(x)\equiv \sum\limits_{Q'}c^{Q'}F^{QQ'}(x),
\quad i=1\div 6
\end{equation}
These forces incorporate in full the symmetric, radial
fluctuations caused by the randomness of the central ball, while all
angular correlations are averaged over and the forces have the full
hexagonal symmetry.
\item
The contact forces between the touching corona balls are assumed to be
given by the effective force law $\tilde{f}(x)$.
\end{itemize}
The displacements of the corona balls are then also symmetric.
This restores the basic picture of a  symmetric
inclusion in the effective lattice. The central
ball remains at its site, while the corona breathes around it symmetrically
in accordance with the central site occupancy,
and transfers this to an equilibrium symmetric distortion of the
surrounding effective lattice.

Returning to Fig.  \ref{fig:3}, we see that the cluster is
surrounded by 12 effective balls
forming a hexagon. They are of two kinds : six occupy the corners ("on top"
positions with respect to the corona balls), and another six sit along the
edges ("bridge" positions). The displacement field around the "1" ligand is
sketched in  Fig. \ref{fig:4}. The scalar $u$ is the radial displacement
common to all corona balls, $u_{\alpha}$ and $u_{\beta}$ correspond to the
on-top and bridge neighbors, respectively. The arrows indicate the
directions of the displacements, whose magnitude varies with the $Q$ label
of the central ball, and should be labelled $u^Q,\, u^Q_1,\,
\ldots$, {\em
etc.}.

Ball 1 is acted upon by contact forces from all its neigbors.
The force coming from the central ball is random (Eq.  (\ref{eq:28})):
\begin{equation}
\label{eq:29}
{\bf
f}_{01}=F^Q(a+u^Q)\,{\hat{\bf n}}_1
\end{equation}
In the regime of small displacements ($\sim$ high pressures), we may expand
\begin{equation}
\label{eq:30}
{\bf
f}_{01}=(F^Q(a)-K^Q(a)u^Q)\,{\hat{\bf n}}_1,\quad K^Q(x)=-{{\rm d}
\over
{{\rm d}x}}F^Q(x)
\end{equation}
The remaining five forces are given by the
effective interaction $\tilde{f}$. They are of three types. In the regime
of small displacements, we obtain ({\em cf.} Eq.  (\ref{eq:5})):
\begin{equation}
\label{eq:31}
\begin{array}{rcl}
{\bf f}_{\alpha 1}&=&\tilde{f}(a)\,{\hat{\bf n}}_4
-\tilde{k}(a)(u^Q-u^Q_{\alpha}) \,{\hat{\bf n}}_4\\
{\bf f}_{\beta 1}&=&\tilde{f}(a)\,{\hat{\bf n}}_5
-\tilde{k}(a)\cdot\\
&&\
\left(\{u^Q\,{\hat{\bf n}}_1-u^Q_{\beta}{1 \over {\surd 3}}
({\hat{\bf n}}_1 + {\hat{\bf n}}_2)\}\cdot{\hat{\bf
n}}_2\right)\,{\hat{\bf n}}_2\\
{\bf f}_{21}&=&\tilde{f}(a)\,{\hat{\bf
n}}_3-\tilde{k}(a)u^Q\,{\hat{\bf n}}_3\\[0.5ex]
&&etc.
\end{array}
\end{equation}
The forces depend on the three displacements
$u^Q, u^Q_{\alpha}, u^Q_{\beta}$. In the linear elasticity
regime corresponding to our assumption of small displacements, a simple
universal relation exists between these three displacements:
\begin{equation}
\label{eq:32}
u_{\alpha}=\alpha\,u,\qquad
u_{\beta}=\beta\,u
\end{equation}
where $\alpha$ and $\beta$ are geometrical parameters independent of the
magnitude of $u$. They can be determined for a linear elastic array
of non-frictional balls once for
ever, although not in a closed analytic form.
Details of the calculation of
both parameters are in the Appendix. The resulting values are
\begin{center}
\bigskip
\begin{tabular}{|c|c|}\hline
$\quad\alpha\quad$
&0.585405\\ \hline
$\beta$&0.232971\\
\hline
\end{tabular}
\end{center}
\bigskip
Now, we require that {\em the total equilibrium force on ball 1 vanishes},
and subtract from this condition the equilibrium condition for the
effective lattice:
\begin{equation}
\label{eq:33}
\begin{array}{rcl}
{\bf
f}_{01}+{\bf f}_{61}+{\bf f}_{\beta 1}+{\bf f}_{\alpha 1}+
{\bf f}_{\beta^{\prime} 1}+ {\bf f}_{21}&=&0\\
\tilde{f}(a)\cdot\{{\hat{\bf n}}_1+{\hat{\bf n}}_{2}+{\hat{\bf
n}}_3+{\hat{\bf n}}_4+ {\hat{\bf n}}_5+{\hat{\bf n}_6\}}&=&0
\end{array}
\end{equation}
Only the component
parallel to the 01 connecting line is non-trivial. With the help of Eq.
(\ref{eq:31}) it reduces to a single equation for a single
unknown, the corona radial displacement, with the
solution
\begin{equation}
\label{eq:34}
u^Q={{\,F^Q(a)-\tilde{f}(a)\,}
\over
{\,K^Q(a)+\Xi\,\tilde{k}(a)\,}}
\end{equation}
where
\begin{equation}
\label
{eq:35}
\Xi={5 \over 2} -\alpha -\beta\,{{\sqrt{3}} \over
2}
\end{equation}
is again a geometrical factor. The equation has a simple interpretation.
For a given $Q$, the difference between the force exerted by the inclusion
and the average force gives rise to a radial displacement.  Its magnitude is
related to the force by an {\em additively renormalized stiffness}
reflecting the fact that the corona is supported by the rest of
the (effective) lattice.  In the mean field context, this renormalization is
a {\em local field correction}, and the parameter $\Xi$ measures its
dimensionless strength.

Now, we may impose the {\em self-consistency condition}, requiring that the
average displacement be zero:
\begin{equation}
\label{eq:36}
\langle u^Q\rangle=\sum\limits_Q c^Q {{F^Q-\tilde{f}}
\over {K^Q+\Xi\tilde{k}}}=0
\end{equation}
where the argument of all functions is $a$.
This equation can be rearranged by introducing:
\begin{equation}
\label{eq:37}
\begin{array}{rcl}
F_\SSC{ave}&=&\langle F^Q\rangle=
\sum\limits_Q c^Q {{F^Q}}\\
{\displaystyle {1 \over {K^{\star}}}}&=&
{\displaystyle \sum\limits_Q c^Q {1 \over {K^Q+\Xi\tilde{k}}} }
\end{array}
\end{equation}
The first quantity is easily recognized as identical with the averaged
force of Eq. (\ref{eq:25}) of the AFA in the
preceding subsection. The second quantity is an averaged radial stiffness
of the lattice surrounding a single ball inclusion. The EMA condition
(\ref{eq:35}) is equivalent to
\begin{equation}
\label{eq:38}
\tilde{f}=F_\SSC{ave}+{K^{\star}}\sum\limits_Q
c^Q
\left({1 \over {K^Q+\Xi\tilde{k}}}-{1 \over {K^{\star}}}\right)
(F^Q-F_\SSC{ave})
\end{equation}
The average hydrostatic force for a given
lattice spacing is seen to contain two contributions. First, the averaged
contact force, and, second, a term having the characteristic structure of a
second-order correlator, known from
CPA-like theories. The second term is also responsible for the peculiar
self-consistent character of the equation (\ref{eq:38}).  It is not
the unknown  $\tilde{f}$ itself which appears on its r.h.s., but
its derivative
$\tilde{k}(a) =-{{\rm d} \over {{\rm d}a}} \tilde{f}(a)$,
and (\ref{eq:38}) has the form
\begin{equation}
\label{eq:39}
\tilde{f}(a)=F_\SSC{ave}(a)+{\cal
F}(\Xi\,{|}\,a,
{\textstyle{{\rm d} \over {{\rm d}a}}}
\tilde{f}(a))
\end{equation}
where ${\cal F}$ is a complicated but well
defined function. The EMA equation is thus  a first order differential
equation determining $\tilde{f}$ as a function of $a$. In other words, we meet
here  a case where  self-consistency cannot be written directly for  an
isolated value of the control parameter $a$, but involves intrinsically the
whole functional dependence $\tilde{f}(a)$.

To select the physically relevant solution of (\ref{eq:39}),
we choose  the {\em high pressure asymptotic boundary condition}
\begin{equation}
\label{eq:40}
\tilde{f}(a)\sim
F_\SSC{ave}(a)\quad\mbox{for}\quad {\mit \Delta}_\SSC{rms}\ll
d-a\ (\,\ll d)
\end{equation}
It can be explicitly verified from(\ref{eq:38})
that the contact of $\tilde{f}$ with $F_\SSC{ave}$ stipulated
by (\ref{eq:40}) is of second order, so that  the boundary
condition is formally justified.  Physically, this condition means
that the EMA coincides with the AFA in the asymptotic limit of high pressures,
but extends the mean field description of the system
to an intermediate pressure region. We may also say that our EMA is a
self-consistent resummation of the high pressure perturbation expansion for
the equation of state $\tilde{f}(a)$.

The solution of the EMA equation (\ref{eq:39}) is obtained without
difficulty by numerical
iteration starting from $\tilde{f}^{(0)}(a)= F_\SSC{ave}(a)$.

\subsubsection{EMA for frictional balls}

The case of frictional balls is very important for a proper comparison with
existing experimental data.
A formal development of the EMA for frictional balls can proceed  quickly,
because the steps are almost identical with those made for the
non-frictional case. It is necessary, however, to extend the description of
the effective medium. As before, it is assumed to be composed of effective
balls with diameter $a$, interacting through effective contacts.
In addition to the average
hydrostatic force $\tilde{f}$ and the normal stiffness
$\tilde{k}=-{{\rm d} \over {{\rm d}a}} \tilde{f}(a)$,
we have to postulate also a local shear (tangential)
stiffness $\tilde{\kappa}$
as a third basic characteristic of the effective contact.
For $\tilde{\kappa}$, we formulate our basic {\em conjecture}:
\begin{equation}
\label{eq:41}
\tilde{\kappa} = \eta\,
\tilde{k}
\end{equation}
In words, the effective shear and compressive stiffnesses are related in
the same manner as the two stiffnesses for actual Hertz-Mindlin contacts
(see Eq.(\ref{eq:6})).
We will discuss this intuitively appealing conjecture below,
but first derive the EMA equations following the preceding paragraph
step by step.

The central ball inclusion and its symmetrized corona are introduced
without changes, and both Figures \ref{fig:3} and \ref{fig:4}
remain valid.  Because of the symmetry of the
cluster, there is no torque acting on either the central or the corona
balls, and only the force equilibrium has to be considered. The central
ball is equilibrated automatically, and we inspect the forces acting on the
corona ball 1,  linearized again with respect to the small displacements.
The full expression (\ref{eq:5}) must be used now. The relative
displacements remain radial for all the neighbors of ball 1, except for the
$\beta$ and $\beta^{\prime}$ ones (see Figure \ref{fig:4}. For
$\beta\,\!$s, the contact force has a shear component,
\begin{equation}
\label{eq:42}
\begin{array}{rclcl}
{\bf f}_{\beta 1}&=&\tilde{f}(a)\,{\hat{\bf n}}_5 \\
&&-\tilde{k}(a)\left(\{u^Q\,{\hat{\bf n}}_1-u^Q_{\beta}{1 \over
{\surd 3}} ({\hat{\bf n}}_1 + {\hat{\bf n}}_2)\}\cdot{\hat{\bf
n}}_2\right)\,{\hat{\bf
n}}_2\\
&&-\tilde{\kappa}(a)\left(\{u^Q\,{\hat{\bf n}}_1-u^Q_{\beta}{1 \over {\surd 3}}
({\hat{\bf n}}_1 + {\hat{\bf n}}_2)\}\cdot{\hat{\bf t}}_2\right)\,
{\hat{\bf t}}_2
\end{array}
\end{equation}
and similarly for $\beta'$. Finally, the displacements
$u^Q, u^Q_{\alpha}, u^Q_{\beta}$
obey, in the linear elasticity regime,
universal relations similar to (\ref{eq:32}),
\begin{equation}
\label{eq:43}
u_{\alpha}=\alpha(\eta)\,u,\qquad u_{\beta}=\beta(\eta)\,u
\end{equation}
where the $\alpha$ and $\beta$ parameters
depend now on the stiffness ratio $\eta$, as indicated.

Introducing Eqs. (\ref{eq:41}) -- (\ref{eq:43}) into the equilibrium
condition (\ref{eq:33}), which is fully general, we obtain the
corona radial displacement as
\begin{equation}
\label{eq:44}
u^Q={{\,F^Q(a)-\tilde{f}(a)\,} \over
{\,K^Q(a)+\Xi_{\eta}\,\tilde{k}(a)\,}}
\end{equation}
in complete analogy with the frictionless case, Eq. (\ref{eq:34}).
The only change is an $\eta$-dependent geometrical factor, which reads
now
\begin{equation}
\label{eq:45}
\begin{array}{rcr}
\Xi_{\eta}&=&{5 \over
2} -\alpha(\eta) -\beta(\eta)\,{{\sqrt{3}} \over 2}\,\\
    &&+\,\eta\,({\,{3 \over 2}\,}  -\beta(\eta)\,{{\sqrt{3}} \over 2})
\end{array}
\end{equation}

The EMA condition $\langle u^Q\rangle=0$, Eq.  (\ref{eq:36}), has a universal
character, and with $u^Q$  given by
(\ref{eq:44}), it can be brought successively to various explicit forms
following strictly Eqs.(\ref{eq:36}) -- (\ref{eq:39}),
with $\Xi$ of Eq.(\ref{eq:35}) replaced everywhere by $\Xi_{\eta}$
as given by (\ref{eq:45}). For convenience, we present here the final form
of the EMA,
\begin{equation}
\label{eq:46}
\tilde{f}(a)=F_\SSC{ave}(a)+{\cal F}(\Xi_{\eta}\,{|}\,a,
{\textstyle{{\rm d} \over {{\rm d}a}}}
\tilde{f}(a))
\end{equation}
The corresponding boundary condition (\ref{eq:40}) is $\eta$-independent.
The two complementary cases, frictional and non-frictional, are thus given
by the same differential equation with the same boundary condition. They
are only distinguished by the value of the $\Xi_{\eta}$ parameter, which
corresponds to $\eta\neq 0$, and $\eta = 0$, respectively. In other words,
the difference between the two physical situations is reflected in the EMA
through the strength of the local field correction appearing in the 
renormalized contact stiffness.

\subsection{Macroscopic properties in the EMA}

To complete the EMA analysis of our granular system, let us now
derive the expressions for
the sound velocities. Macroscopically, the averaged system
is a periodic
h.c.p. structure, so that it is acoustically isotropic and there exist just
two sound velocities $v_L$ and $v_T$, as in the ideal lattice of Sec. IIB.
There, we proceeded from the full dispersion law in the Brillouin zone to
the long wave limit (Eqs.(\ref{eq:13}), (\ref{eq:14})), then to the elastic
moduli (\ref{eq:17}) on the basis of the macroscopic (continuum) relations
(\ref{eq:16}).

For the disordered system, we shall go in the reverse direction. The EMA
yields directly the equation of state $\tilde{f}(a)$ and the effective
compressive stiffness $\tilde{k}(a)$. The equation of state is
a monotonous function of $a$ and thus can be inverted into $\tilde{a}(f)$.
Using Eq. (\ref{eq:20})), we obtain finally the average lattice spacing as a
function of the linear pressure $P$, and any function of $a$, such as the
bulk modulus, can be converted into a function of the pressure.

The bulk modulus is obtained from the stiffness $\tilde{k}$
using Eq. (\ref{eq:21}). It should be stressed that Eqs. (\ref{eq:17})
and (\ref{eq:21}) have an identical form; in fact, (\ref{eq:17}) as far as
it concerns the bulk modulus is but a special case of (\ref{eq:21}) for
zero disorder.  As concerns the shear modulus, since our method only allows for
treating the effect of hydrostatic stresses, we resort to an {\em ansatz}, 
i.e., we propose to extend
relation (\ref{eq:17}) for $G$ to the disordered case. We then get :
\begin{equation}
\label{eq:47}
K_\SSC{eff} = \frac{\sqrt{3}}{2}\frac{\tilde{k}}{d},\qquad
G_\SSC{eff} = \frac{(1 + \eta)}{2}\,K_\SSC{eff}
\end{equation}
While the first relation is exact, we should discuss the meaning of the
second one. It turns out that the equivalent relation (\ref{eq:17}) for the
ideal array is a direct consequence of the Mindlin ratio (\ref{eq:6})
between near-normal stiffnesses and does not depend on the specific value
or character of the compressive stiffness. Thus, the second of Eqs.
(\ref{eq:47}) is in fact equivalent to our conjecture
(\ref{eq:41}) about the ratio of effective stiffnesses,
$\tilde{\kappa} = \eta\, \tilde{k}$. Once the effective moduli are known,
the sound velocities become
\begin {equation}
\label{eq:48}
v_{L}^{2} =
\frac{K_\SSC{eff} +G_\SSC{eff}}{\tilde\rho},\qquad
v_{T}^{2} =
\frac{G_\SSC{eff}}{\tilde\rho}
\end{equation}
Equations (\ref{eq:47}) then predict for the velocity ratio that
\begin{equation}
\label{eq:49}
{{v_{L}^{2}} \over
{v_{T}^{2}}}={{\,3+\eta\,} \over {1+\eta}}
\end{equation}
This is a very strong prediction, according to which the velocity ratio
should depend neither on the pressure nor on the disorder level,
being fully determined by the material properties of the grains.
While conjecture (\ref{eq:41}) is not susceptible of direct verification, the
predicted sound velocity ratio can be tested for a wide range of systems.
Note that a similar conjecture concerning this ratio was formulated earlier
by Schwartz and Johnson \cite{Fenghcp}.


\section{Discussion}
We first confront our mean field predictions with the
experimental data obtained by Gilles and Coste \cite{Gilles}, obtained on
an h.c.p. lattice of steel balls confined by a
hexagonal frame under hydrostatic loading. They measured the time of
flight of a low frequency sound pulse between two opposite sides of the
frame. From this we obtain the longitudinal sound velocity $v_{L}(f)$,
where $f$ is the average equilibrium contact force (Eq.(\ref{eq:20}).

The experimental data show that, at the highest loadings used, $v_{L}$
approaches closely the $P^{1/6}$ Hertz behavior. This confirms that, in
this regime, the applied pressure is high enough for disorder effects to be
weak, and for the compressed lattice to be close to the ideal h.c.p.  one.

We therefore focus first on
the data for the largest forces used in
the experiments. 
For example, an external force of $\sim 990\,$N applied
on each
side of a hexagon with 31 balls in contact with each side
corresponds to
 $f = 18.446\,$N. Taking for
steel $E = 9.20\, 10^{10}\,$Pa,
$\sigma = 0.276$, $\rho = 7840\,$%
kg.m$^{-3}$, ball diameters $d = 8$mm,
we get from Eqs. (\ref{eq:2}),
(\ref{eq:3}) and
(\ref{eq:13})  $v_{L} =
791\,$m.sec$^{-1}$, to be compared
with the
experimental value of
$778\,$m.sec$^{-1}$. The agreement is seen to be 
excellent,
to the
precision of the numerical input. 

Note that the contribution of the
tangential contact stiffness
is non negligible : with the above value of
$\sigma$,
the Mindlin coefficient $\eta = 0.84$. Neglecting the
frictional
character of the contacts ($\eta = 0$) leads to underestimating
the
longitudinal
velocity by $\approx 11\%$,
well out of the experimental
error bar.
For the transverse velocity, the two values would differ by
$\approx 35\%$,
but no experimental data are available. These differences
in velocities are
clearly reflected in the different slopes of the
dispersion curves for the
frictional and non-frictional cases in Fig.
\ref{fig:2}.

With the values of $v_L$ at hand, we may return to the
discussion at the
end of Sec. IIB and check first the validity of the Hertz
approach. The
bulk sound velocities for the above material parameters of
steel are
$v_{L,\SSC{bulk}} = 5825\,$m.sec$^{-1}$,
$v_{T,\SSC{bulk}} =
3252\,$m.sec$^{-1}$ and we may conclude that the
condition $v_{L,T}\ll
v_{L,T,\SSC{bulk}}$ is obeyed.
To verify the validity of the continuum
limit, we have to invoke the
ground frequency of  the pulses used in the
experiments, 
$\omega = 2\pi\times 6\,500\,$Hz. This yields
$q\cdot(2\pi/d)^{-1} \approx 0.066$, much less than 1 indeed. 
Expressed in terms of wavelengths,
$\lambda \approx 0.122\,{\rm m} \sim 15 d$. These estimates made for one
pressure may be taken as representative for the whole experimental pressure
range, as the experimental velocities vary between 500 and 
$800\,{\rm m.sec}^{-1}$.

In order to analyze the pressure dependence, we deduce from the
$v_{L}$ data the effective normal stiffness $\tilde{k}$
with the help of equations (\ref{eq:49}), (\ref{eq:48}),
assuming frictional contacts with $\eta = 0.84$.
It is plotted on Figure \ref{fig:5}.
It is clearly seen that the high pressure data fit the straight line
corresponding to the ideal lattice. As $f$ decreases, the logarithmic
slope increases, coming close to the popular $\nu\approx 1/4$ value.
However, no sharp crossover is identifiable, the transition being
completely smooth.

We then solve the mean field equation numerically. All
parameters are known, except for the width of the distribution of ball
diameters. Diameter scatter is only qualified, in the experiment, through a
tolerance of $\pm 4\,\mu$m. We assume a uniform distribution whose width
$W$ is our single fitting parameter. We find that the best fit is obtained
for $W = 2.04\,\mu$m, 
compatible with the tolerance figure. The fit itself is seen to be very
good in the whole experimental range.  The EMA is shown both for frictional
and non-frictional balls. The results are only weakly sensitive to the
value of $\eta$, at  striking variance with the sound velocity itself. For
the latter quantity, however, the $\eta$ dependence enters primarily
through the $G_\SSC{eff}/K_\SSC{eff}$ ratio.

The experimental $f$-range is
limited, on the low force side, to $f \simeq 2$N.
Furthermore, no direct measurement of the equation of state $\tilde{f}(a)$ is
available yet. It is therefore of great interest to complement the above
comparison with a confrontation between mean field predictions and the
simulations performed by Roux\cite{Roux1} on the same system for
frictionless balls with a uniform random distribution of diameters. Applied
forces span the whole range from almost zero up to the upper experimental
limit.  The comparison for the equation of state is shown on Figure
\ref{fig:6}, which is drawn in the dimensionless representation used in
reference \cite{Roux1}:
$$
a^{*}={{d+{\textstyle {1 \over 2}}W - a}
\over W},\quad
f^{*}={ f \over {
{\textstyle {2 \over 3}}E^* d^{1 \over
2}
W^{3 \over 2} }}
$$
In these units, the exact equation of state $f^*(a^*)$ is a universal function
\cite{Roux1}, if approximation (\ref{eq:22}) is used. 
It is easy to verify on Eq. (\ref{eq:38}) that the EMA has the same property.

Clearly, $a^{*}\ge 0$. For $a^{*}=0$, the balls barely start touching,
$a_{o}^{*} = .343$ is the so-called rigidity threshold,
below which the disordered lattice cannot sustain compression.
$a^{*} \ge 1$ is the region where all neighbors are already in contact,
and $a^{*} \gg 1$ is the high pressure limit.  In a narrow range
above the rigidity threshold, a quasi-Hertz regime is found, followed, in
the intermediate force range, by a steeper variation. The mean field result
shown here, also calculated for frictionless
balls, appears to agree with the numerical data for
$(a^{*}-a_{o}^{*}) \gtrsim  0.3$. Also shown are the curves corresponding 
to the ALA and to the AFA. They frame the exact and the EMA results from 
below and from above, respectively. For high pressures, all plots merge.
We replot the same data in Fig. \ref{fig:7} using the linear
scale for the lattice spacing, and the $2 \over 3$ power scale for the
forces, so that the natural range of various regimes can be
better assessed. In particular, note the striking
improvement of EMA over the AFA for intermediate pressures.

In Fig.  \ref{fig:8} we return to the experimentally more relevant
 stiffness-force  dependence and plot in dimensionless form
the mean-field and the ALA  curves, the data from Roux's
numerical simulations and their power law approximants, and also the
experimental data of Gilles and Coste.  These data are scaled using $W =
2.04\,\mu$m and the force unit 
${{2 \over 3}}E^* d^{1 \over 2} W^{3 \over 2} = 19.89\,$N.
The overall agreement is truly satisfactory, which is all
the more remarkable that, while experiments and mean field are concerned
with frictional balls, the simulations relate to the frictionless case.
This is to be related with our previous observation that, in the h.c.p.
lattice studied here, the normal stiffness is only weakly sensitive to
shear interball forces. Such might not be the case for other ball arrays.

As is well known, since mean field theories are not systematic
expansions in powers of a small parameter, they do not allow for a
precise direct assessment of their range of validity.  From this
discussion, we can state empirically that the validity of our effective
medium approximation is limited to dimensionless forces $f^{*}\gtrsim
0.08$. This we confirm by calculating, within the mean field approximation
itself, the relative force fluctuations 
${\mit \Delta} f^{*}/f^{*} = {\mit \Delta} \tilde{f}/\tilde{f}$,
and ${\mit \Delta} F_\SSC{ave}/ F_\SSC{ave}$, where
\begin{equation}
\label{eq:fluctuation}
\begin{array}{rcl}
({\mit \Delta}
f)^{2} &=& \sum\limits_{Q}\,c_{Q} (F^{Q}(a) - K^{Q}u^{Q} - \tilde{f}(a))^{2})
\end{array}
\end{equation}

The plot of Figure  \ref{fig:9} shows that the relative fluctuations
decrease rapidly with increasing displacement, i.e. pressure, the 
above mentioned empirical limit corresponding to the reasonable value
${\mit \Delta} f^{*}/f^{*} \simeq 0.6$. The reduction of the local stress (force)
fluctuations due to the self-consistent local displacements is reflected 
in the relative magnitude
of the EMA and AFA fluctuations. The frictionless balls appear to adjust
their positions more effectively, as could be expected.

Finally, we have estimated the average fraction of active contacts per ball
$N_{c}$ in our effective lattice in the following way. We define it
to be the average number of neighbors with an intercenter distance smaller 
than the sum of the corresponding radii.
\begin{equation}
\label{eq:number}
N_{c} =
\sum\limits_{Q,Q'}\,c_{Q}c_{Q'}\,\vartheta
(\frac{d_{Q}+d_{Q'}}{2} - a -
u_{Q})
\end{equation}
Note that this expression does not derive
systematically from the mean field formalism, in which the notion of
contact number does not enter explicitly. It should therefore be considered
as indicative only. $N_{c}$ is plotted on Figure \ref{fig:10} together
with the values calculated by Roux. Expression (\ref{eq:number}) is seen
to systematically underestimate $N_{c}$. It appears that the validity
of the mean field extends down to $N_{c}\sim 85\%$. However, it is
important to point out that the pressure range where the sound velocity
departs from the Hertz behavior extends well into the pressure range where
the contact fraction has already saturated to $1$.  This corroborates
strongly the idea that it is not the connectivity itself which is
responsible for the non-Hertz behavior, but the presence of
disorder-induced stress fluctuations. This is shown in Fig. \ref{fig:11},
where we plot together the EMA results for $N_c$,
${\mit \Delta} f^{*}/f^{*}$ and the ratio $\tilde{k}/k_\SSC{ala}$. 
This latter quantity
is a direct measure of the deviations of the effective stiffness from the
ideal Hertz law. No sharp change of regime occurs at saturation of $N_c$,
the non-Hertz behavior and stress fluctuations appear to
extend to higher pressures and gradually tend to zero together.

\section{Conclusion}
On the basis of the above discussion, we are therefore able to
conclude that our mean field theory provides quite a satisfactory
description of the pressure dependence of the bulk mechanical
properties, and hence of the sound velocity, in an array of
frictional balls in the "high" pressure range. This corresponds to
the regime in which the ball network is strongly overdetermined, that
is where connectivity is close to its saturation value.

We believe that this agreement, which permits to account for the
non-Hertzian behavior down to the $\nu \approx 1/4$ range, is due to
the fact that this theory does capture, though in an approximate
manner, the existence of disorder induced stress fluctuations, and
that these are self-consistently related with the global
mechanical state of the system. In other words, the true disorder
strength in the problem is not intrinsically given by the dispersion
of unstressed ball diameters, but determined from this $together\,with$ the
elastic deformation field.

We have based our single-site description upon the simplest possible
approximation, which amounts to a spherical averaging of local
lattice distorsions. That this is sufficient to produce a
satisfactory theory must certainly be attributed to the high
connectivity of the h.c.p. structure. It should for this reason be
applicable as well to a 3D lattice such as the fcc one.

Systematic improvement upon this approximation is possible. This
should be based on extending the basic building block from our
(central ball + average bond star) to a full cluster made of the
central ball and of its six neighbors.  Such an extension, although
obviously heavy, would have the merit of explicitly allowing for local
asymmetric configurations, which we have ignored here. It would also
be a first step towards taking into account spatial stress
correlations, which are completely overlooked in the present approach.

In view of this last remark, the good agreement obtained here with
sound velocity data calls for a physical comment. Coherent sound
measures a mechanical response on the scale of the corresponding
effective wavelength which, in the low frequency regime of the
Gilles-Coste experiments, is at least of the order of 10 ball
diameters, as discussed in Sec. IV.  It is now well documented
\cite{Radjai}, \cite{Jia}, \cite{Miller} that the correlation length of the
stress network in granular packings is, at most, of order a few $d$. It is
therefore to be expected that mean field theories, though by nature unable
to capture long range correlations, are well adapted to describe large
scale properties. That is, when focussing on large scale mechanical
responses, the pertinent notion is that of a global fluctuating stress
network. The complementary notion of stress chains, which emphasizes the
long range part of correlations, is certainly more relevant to the
question of acoustic scattering, which becomes essential at higher
frequencies.

Obviously, an important pending question is concerned with the
possibility of extending the mean field approach to the more
important case of topologically disordered random grain packings.
Such a step, however desirable it might be, is by no means straightforward, as
can be inferred from the, yet insuperable, difficulties which have been met
when trying to extend the theories of electronic and vibrational properties
of random substitutional alloys to amorphous materials. These can be
assigned to the absence of a natural reference configuration possessing
long range order. Up to now, the only existing theoretical frame for
amorphous solids is a structureless, homogeneous effective medium. This
would amount in the present problem to treating the effective medium in
the continuum limit, which is clearly inedequate to account for stress
fluctuation effects. In this perspective, numerical studies appear
essential as a basis for trying to build up the needed original theoretical
concepts.

However, we believe that the qualitative idea which emerges from this
work, namely that it is the disorder induced stress fluctuations which
are responsible for the pressure dependence of the sound velocity in
granular packings, will carry over, as well, to topologically
disordered systems.

\acknowledgements
We are grateful to C.  Coste, B. Gilles and J.N. Roux for fruitful
discussions and for communication of their results prior to publication.
B.V. gratefully acknowledges the hospitality of Universit\'e Paris VII.
This work was also supported in part by the Grant Agency of the Charles
University of Prague (project 146/1999).

\appendix
\section{Calculation of \protect{\boldmath $\alpha,\,\beta$}}
We want here to calculate the expression of the coefficients
$\alpha$, $\beta$ defined in Section III. For this purpose, we need
to solve the following problem. Consider an ideal h.c.p. lattice,
with interball distance at equilibrium $d$, in which we singularize
a central site $(0)$. Apply to the six balls $i = 1,\ldots,6$ of the
corona of its nearest neighbors excess forces directed along the $(0i)$
bonds ${\bf F}_{i} = F\hat{\bf n}_{i}$ (see Figure \ref{fig:4}).
Call $ {\bf u}_{i} = u_{i}\hat{\bf n}_{i}$ the resulting displacement 
of nearest neighbor $(i)$.
Then, by definition, $\alpha  = u_{i\alpha}/u_{i}$, where
${\bf u}_{i\alpha}$ is the displacement of the second neighbor
$(i\alpha)$ along the $\hat{\bf n}_{i}$ direction;
$\beta = u_{i\beta}/u_{i}$, with ${\bf u}_{i\beta}$ the
displacement of the next nearest neighbor along the direction of
$\frac{1}{\surd{3}}{\hat{\bf n}_{i} + \hat{\bf n}_{i+1}}$.

In order to avoid excessive algebraic heaviness, and in view of the
fact that numerical studies lead us to conclude to the
very weak sensitivity of the mean field results to moderate variations of
$\alpha$, $\beta$, we limit ourselves to the simple case of vanishing
Mindlin shear stiffness: $\eta = 0$.

In this case rotations become irrelevant and, in the linear response
regime, the force on a ball is related to the displacements of its
neighbors by:
\begin{equation}
\label{eq:A1}
{\bf F}_{i} = \sum \limits _{\{j\}}
\overline{\overline{D}}_{ij}.{\bf u}_{j}
\end{equation}
So that, inverting in Fourier space :
\begin{equation}
\label{eq:A2}
{\bf u}_{i} =
\sum \limits _{\bf k}{\rm e}^{i{\bf
q}.{\bf R}_{i}}\,\overline{\overline{\Delta}} ({\bf q})\,{\bf F}({\bf q})
\end{equation}
with
\begin{equation}
\label{eq:A3}
\overline{\overline{\Delta}}({\bf q}) = \overline{\overline{D}}^{-1}({\bf
q})
\end{equation}
From equation(\ref{eq:9}), $\bar{\bar{D}}({\bf q})$ is a $2\,\times \,2$
matrix acting in ($x,y$) space, defined by:
\begin{equation}
\label{eq:A4}
\begin{array}{rcl}
\overline{\overline{D}}({\bf q})&=&
\sum\limits_{i,\{ j\}} {\rm e}^{i{\bf q}.
({\bf R}_{i}-{\bf R}_{j})}\,\overline{\overline{D}}_{ij}\\
&=& 2k\sum\limits_{p=1}^{3}
(1-\cos(d{\bf q}.\hat{\bf n}_{p}))\,\hat{\bf n}_{p}\hat{\bf n}_{p}
\end{array}
\end{equation}
The applied forces are defined by:
\begin{equation}
\label{eq:A5}
{\bf F}_{i} = F\,[\hat{\bf
n}_{1}(\delta_{i1}-\delta_{i4})
+\hat{\bf
n}_{2}(\delta_{i3}-\delta_{i6})+\hat{\bf
n}_{3}(\delta_{i5}-\delta
_{i2})]
\end{equation}
where the unit vectors $\hat{\bf n}_{p}$ and the sites $(i)$ are
labelled according to Figure \ref{fig:1}.

Then:
\begin{equation}
\label{eq:A6}
{\bf F}({\bf q}) =
\frac{f}{N}\sum\limits_{p=1}^{3}\,(-2i)\,\hat{\bf
n}_{p} \sin(d{\bf q}.\hat{\bf n}_{p})
\end{equation}
N being the number of sites in the lattice.

Taking advantage of the fact that, by symmetry, 
$\hat{\bf n}_{1}.({\bf u}_{1}-{\bf u}_{4}) = \hat{\bf n}_{2}.({\bf u}_{3}-{\bf
u}_{6}) = \hat{\bf n}_{3}.({\bf u}_{5}-{\bf u}_{2})$, one gets:
\renewcommand{\arraystretch}{2.}
\begin{equation}
\label{eq:A7}
\begin{array}{r@{}c@{}l}
u_{1}&=&{\displaystyle\frac{\hat{\bf
n}_{1}.({\bf
u}_{1}-{\bf u}_{4}) }{2}}\\&=&{\displaystyle
\frac{f}{3N}\sum\limits_{{\bf
q}}\sum\limits_{p,r=1}^{3}}\\&&\hat{\bf
n}_{p}.\overline{\overline{\Delta}}({\bf
q}).\hat{\bf n}_{r}(\cos[d{\bf
q}.(\hat{\bf n}_{p}-\hat{\bf n}_{r})]-
\cos[d{\bf q}.(\hat{\bf
n}_{p}+\hat{\bf n}_{r})])
\end{array}
\end{equation}
Analogously, the displacement $u_{1\alpha}$ of the second neighbor along $Ox$:
\begin{equation}
\label{eq:A8}
\begin{array}{r@{}c@{}l}
u_{1\alpha}&=&{\displaystyle
\frac{f}{3N}\sum\limits_{{\bf
q}}\sum\limits_{p,r=1}^{3}}\\&&\!\!\!\!\!\hat{\bf
n}_{p}.\overline{\overline{\Delta}}({\bf
q}).\hat{\bf n}_{r}(\cos[d{\bf
q}.(2\hat{\bf n}_{p}-\hat{\bf n}_{r})]-
\cos[d{\bf q}.(2\hat{\bf
n}_{p}+\hat{\bf n}_{r})])
\end{array}
\end{equation}
From equations (\ref{eq:3}),(\ref{eq:4}),
\begin{equation}
\label{eq:A9}
{\cal D}({\bf
q}) = \det[\overline{\overline{D}}({\bf q})] = \sum
\limits_{p=1}^{3}\,
\Gamma_{p}\Gamma_{p+1}
\end{equation}
with
\begin{equation}
\label{eq:A10}
\Gamma_{p} = (1-\cos(d{\bf
q}.\hat{\bf
n}_{p})\,\,\,\,\,\,\,\,\,\,\,\,\,\,\,\,\,\Gamma_{p+3} \equiv
\Gamma_{p}
\end{equation}
Finally, one obtains, for $\alpha = u_{1\alpha}/u_{1}$:
\begin{equation}
\label{eq:A11}
\alpha =
\frac{\sum\limits_{\bf q}\,{\cal
D}^{-1}({\bf
q})\,\sum\limits_{p=1}^{3}\,\Gamma_{p}(\Sigma_{p+1}^{(2)}-
\Sigma_{p+2}^{(2)
}) (\Sigma_{p+1}^{(1)}-
\Sigma_{p+2}^{(1)})}{\sum\limits_{\bf
q}\,{\cal
D}^{-1}({\bf
q})\,\sum\limits_{p=1}^{3}\,\Gamma_{p}(\Sigma_{p+1}^{(1)}-
\Sigma_{p+2}^{(1)
})^{2}}
\end{equation}
where we have set:
\begin{equation}
\label{eq:A12}
\Sigma_{p}^{(n)}=\sin(nd{\bf
q}.{\hat{\bf
n}}_{p})\,\,\,\,\,\,\,\,\,\,\,\,\,\,\,\, T_{p}=\sin(d{\bf q}.{\hat{\bf
t}}_{p})
\end{equation}
A completely analogous calculation involving the displacement
$u_{1\beta}$ of the second nearest neighbor yields:
\begin{equation}
\label{eq:A13}
\beta =\frac{\sum\limits_{\bf
q}\frac{2}{\sqrt
3}{\cal
D}^{-1}({\bf
q})\sum\limits_{p=1}^{3}\,\Gamma_{p}(T_{p}-\frac{T_{p+1
}+T_{p+2}}{2})
(\Sigma_{p+1}^{(1)}-\Sigma_{p+1}^{(2)})}{\sum\limits_{\bf
q}\,{\cal
D}^{-1}({\bf
q})\,\sum\limits_{p=1}^{3}\,\Gamma_{p}(\Sigma_{p+1}^{(1)}-
\Sigma_{p+2}^{(1)
})^{2}}
\end{equation}
Performing numerically the ${\bf q}$-integrations over the Brillouin
zone, we obtain:
\begin{equation}
\label{eq:A14}
\alpha
=0.585405\,\,\,\,\,\,\,\,\,\,\,\,\, \beta =
0.232971
\end{equation}

\newpage
%
\begin{figure}
\caption{
\label{fig:1}
Sketch of the ideal 2D h.c.p.  lattice.
}
\end{figure}

\begin{figure}
\caption{
\label{fig:2}
Dimensionless dispersion curves for vibrations in the h.c.p. lattice
along two principal directions in the Brillouin zone.
Units:$(k/M)^{{1 \over 2}}$ for frequency, $2\pi/d$ for wave vector.
Upper panel: frictional balls, $\eta_\SSC{steel}=0.84$. Lower panel:
frictionless balls. Middle: Brillouin zone, 
$\overline{\Gamma K}= {2 \over 3}{2\pi/d}$, 
$\overline{\Gamma M}= {1 \over {\surd 3}}{2\pi/d}$.
}
\end{figure}

\begin{figure}
\caption{
\label{fig:3}
Graphical representation of the EMA averaging process.
On the left, a cluster (central ball 0 surrounded by its corona, balls
1 $\div$ 6) embedded in the effective medium. Configuration averaging
$\langle\ldots\rangle$ restores the effective medium on the right.
}
\end{figure}

\begin{figure}
\caption{
\label{fig:4}
Displacements of the neighbors of the cluster central ball 0.
Corona balls: 1, 2, 6.  Second neighbors: effective balls in the
on-top ($\alpha$) and bridge ($\beta$ and $\beta'$) positions.
}
\end{figure}

\begin{figure}
\caption{
\label{fig:5}
Square root of  effective compressive stiffness  {\em vs.} effective
contact force.  Squares: experiment (Ref. \protect\cite{Gilles}).
Thick line: EMA for frictional balls. Dashed line: EMA for frictionless balls.
Thin line: ALA (ideal Hertz dependence).
}
\end{figure}

\begin{figure}
\caption{
\label{fig:6}
Effective force {\em vs.} displacement in dimensionless units (see text)
for frictionless balls.
$a^*_o$ corresponds to the rigidity threshold. Dots: numerical results.
Dash-dotted lines: power law approximants of numerical results (Ref.
\protect\cite{Roux1}).
}
\end{figure}

\begin{figure}
\caption{
\label{fig:7}
Effective force {\em vs.} displacement in the "Hertz" representation.
Data and symbols as in Fig.  \ref{fig:6}.
}
\end{figure}

\begin{figure}
\caption{
\label{fig:8}
Square root of  effective compressive stiffness {\em vs.} effective
contact force (dimensionless units).
}
\end{figure}

\begin{figure}
\caption{
\label{fig:9}
Relative fluctuation of local contact force {\em vs.} dimensionless
displacement.  Full line: EMA for frictional balls. Thick dashed line:
 EMA for frictionless balls. Dotted line: AFA.
}
\end{figure}

\begin{figure}
\caption{
\label{fig:10}
Active contact fraction {\em vs.} dimensionless
displacement. Full line: EMA for frictional balls. Thick dashed line:
 EMA for frictionless balls. Dots: numerical simulation \protect\cite{Roux1}.
}
\end{figure}

\begin{figure}
\caption{
\label{fig:11}
The relative departure $k^*/k^*_\SSC{ala}$ of the EMA effective normal
stiffness from ideal Hertz and the EMA force fluctuation 
${\mit \Delta} f^*/f^*$
exhibit no sharp change at the point $f^* \simeq 0.3$ where the active
contact fraction $N_c$ saturates to 1.
}
\end{figure}


\begin{references}

%
\bibitem[*]{Charles} Permanent address : Faculty of Mathematics and
Physics, Charles University, Ke Karlovu 5, 121 16 Praha 2, Czech Republic

%
\bibitem[\dag]{CNRS} ``Associ\'e au Centre National de la
Recherche Scientifique  et aux Universit\'es Paris 6 et Paris
7''.
\bibitem{KLJ} K.L. Johnson, {\it Contact Mechanics}, (Cambridge
Univ. Press, Cambridge, 1985).
\bibitem{Duffy} J. Duffy and R.D.  Mindlin, 
J. Appl. Mech., {\bf 24}, 585 (1957).
\bibitem{Gilles} B. Gilles, and C. Coste, in {\it Powders and Grains 2001}
Y. Kishino, Ed.  (A. A. Balkema, Lisse 2001) {\em p.} 113
\bibitem{Goddard}J.D. Goddard,
Proc. R. Soc Lond. {\bf A430}, 105 (1990).
\bibitem{de Gennes} P.-G. de Gennes, Europhys. Lett. {\bf 35}, 145 (1996).
\bibitem{Makse} H.A.  Makse, N. Gland, D.L. Johnson, and L.M. Schwartz, 
Phys.  Rev. Lett. {\bf 83}, 5070 (1999).
\bibitem{Roux1}J.N. Roux, in {\it Powders and Grains 97}
R. Behringer and J. Jenkins eds.,
(A. A. Balkema, Rotterdam, 1997) {\em p.} 215;\\
J.N. Roux, in  Actes du Colloque {\it Physique et Mecanique
des Milieux Granulaires} (LCPC, Champs sur Marne, 2000).
\bibitem{Digby}
P.J. Digby, J. Appl. Mech., {\bf 48}, 803 (1981).
\bibitem{Walton} K.  Walton, J. Mech. Phys. Solids {\bf 35}, 213 (1987).
\bibitem{Winkler} K.W. Winkler, Geophys. Res. Lett., {\bf 10}, 1073 (1983).
\bibitem{Sheng} P. Sheng, {\it Introduction to Wave Scattering,
Localization and Mesoscopic Phenomena} (Academic Press, San
Diego, 1995).
\bibitem{FengMF1} S. Feng, M.F. Thorpe,  and E. Garboczi,
Phys. Rev. {\bf B31}, 276 (1985).
\bibitem{FengMF2} L.M. Schwartz, S.  Feng, M.F. Thorpe, and P.N. Sen,
Phys. Rev. {\bf B32}, 4607 (1985).
\bibitem{Kirkpatrick}S. Kirkpatrick, Rev. Mod. Phys., {\bf 45}, 574 (1973).
\bibitem{Fenghcp} L.M. Schwartz and D.L. Johnson,
Phys. Rev. Lett.,{\bf 52}, 831 (1984).
\bibitem{plastic} Note that at applied forces where this condition
would not hold, the normal stress within the contact would reach a
level comparable with the yield stress of the material, and plastic
effects would have to be taken into account.
\bibitem{Jia} X. Jia, C. Caroli, and B. Velicky, Phys. Rev.
Lett., {\bf 82}, 1863 (1999).
\bibitem{Radjai} F. Radjai, M. Jean, J.J.  Moreau, and S. Roux, Phys. Rev.
Lett. {\bf 77}, 274 (1996).
\bibitem{plus} We use the shorthand $x^\alpha_{+}$ for a function which is
zero for $x<0$ and equals $x^\alpha$ for $x\ge 0$.
\bibitem{approx} Approximations closely resembling the two ones
described here are known in various areas. Thus, the Averaged lattice
approximation corresponds to the Virtual crystal approximation in alloy
physics and to the Reuss approximation in mechanics, while the Averaged
force approximation lies between the Atomic limit approximation and the
Averaged t-matrix approximation in solid state physics, and is known as the
Voigt approximation in mechanics.
\bibitem{Miller} B. Miller, C.  O'Hern, and R. P. Behringer, Phys. Rev.
Lett. {\bf 77}, 3110 (1996).
%
\end{references}
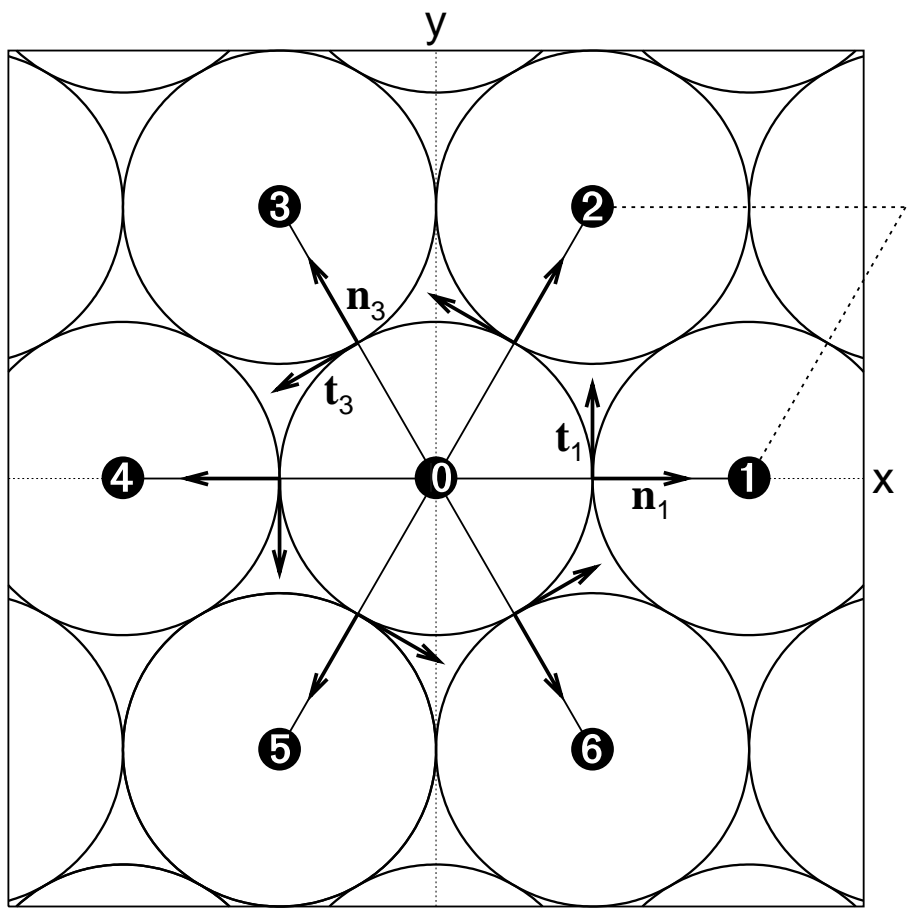
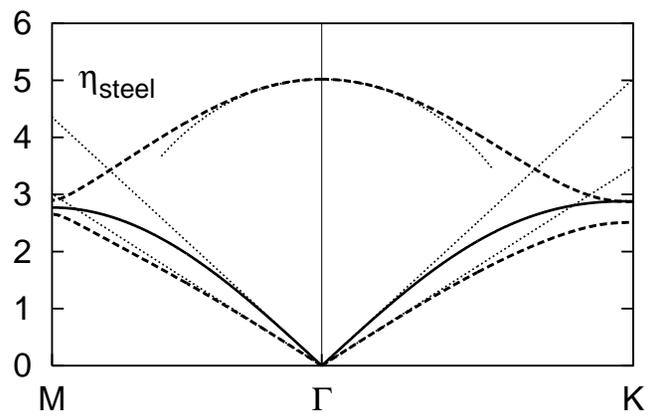
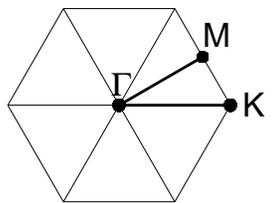
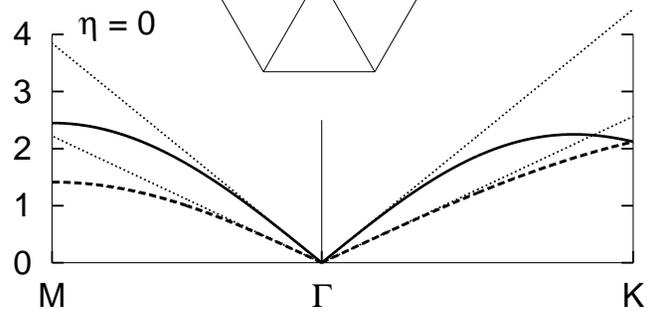
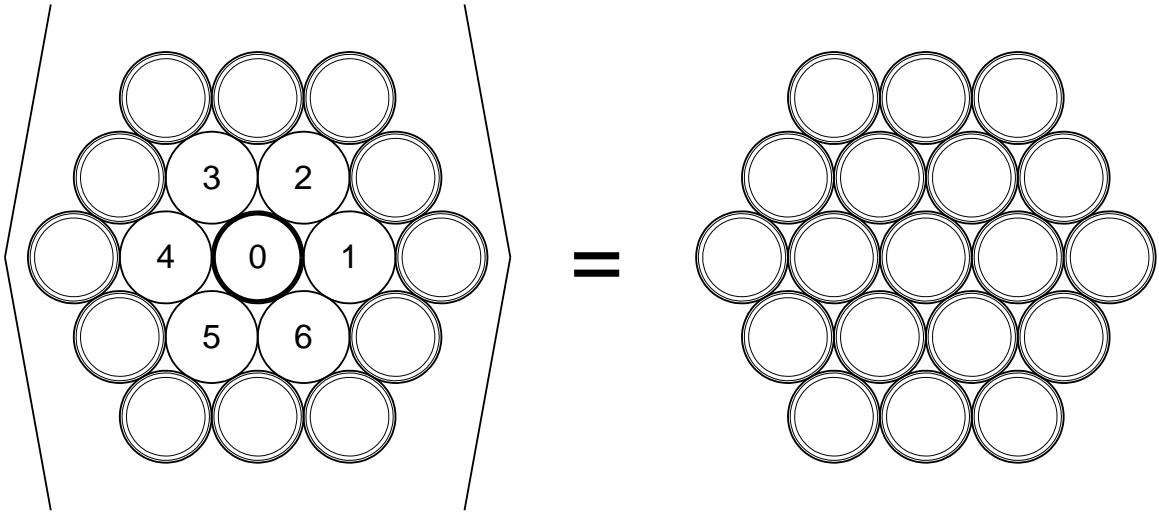
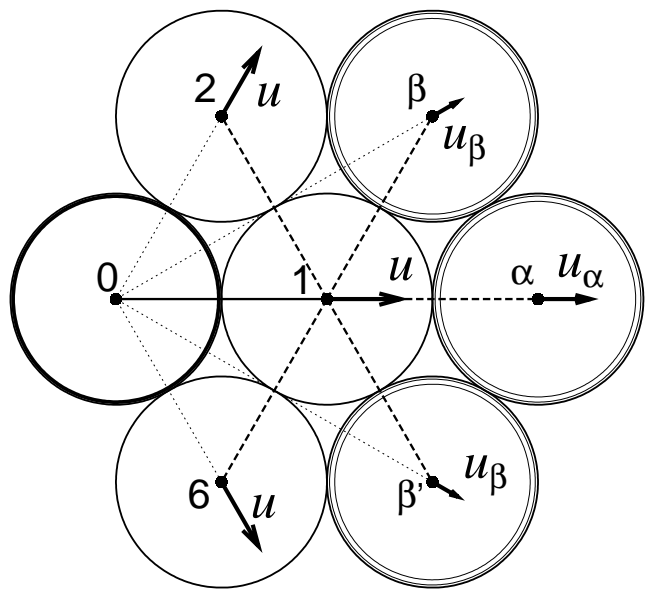
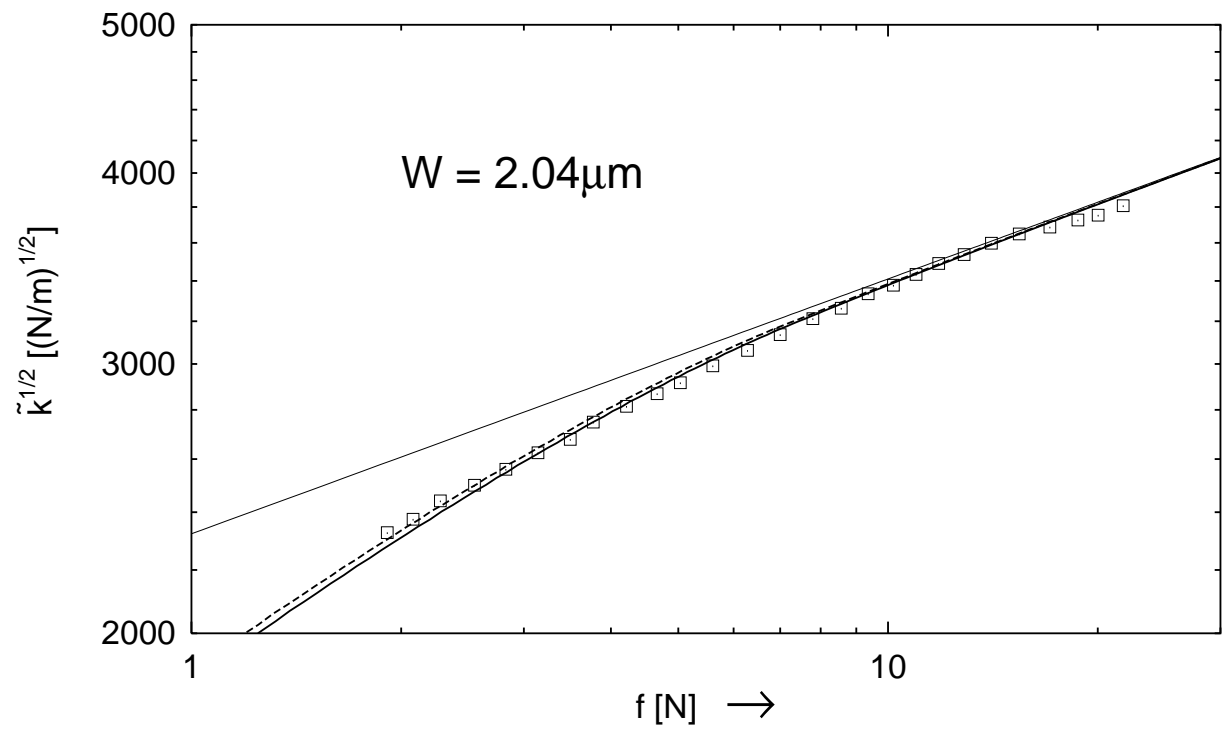
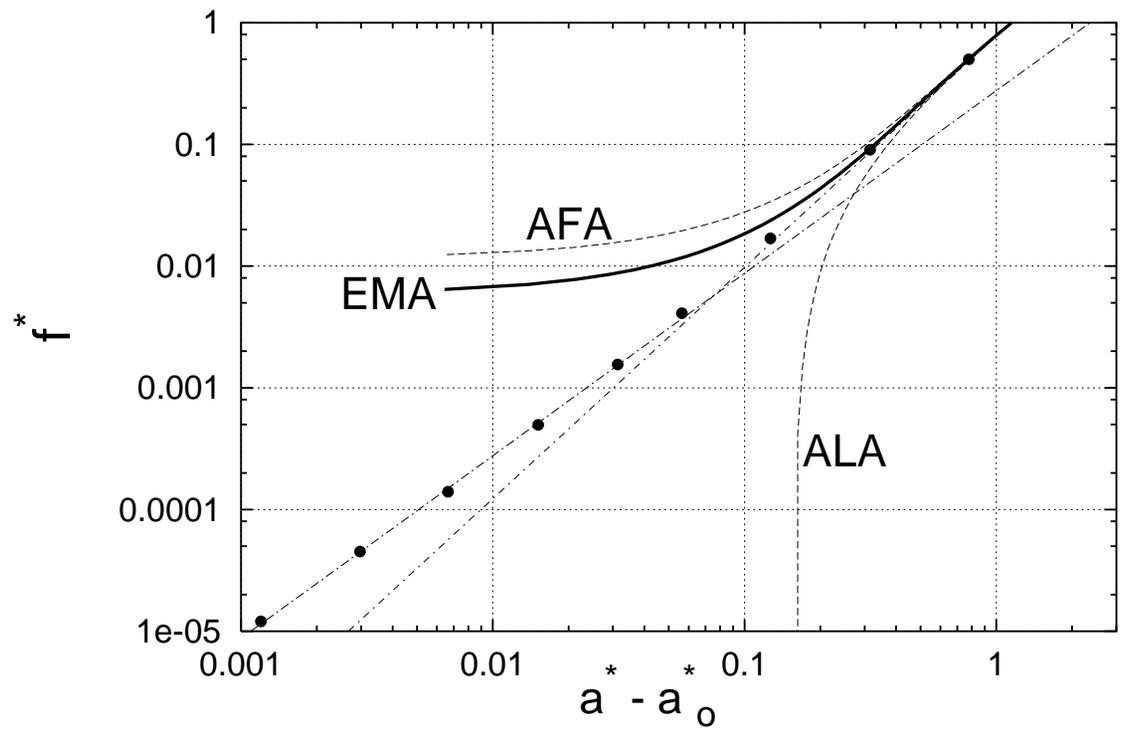
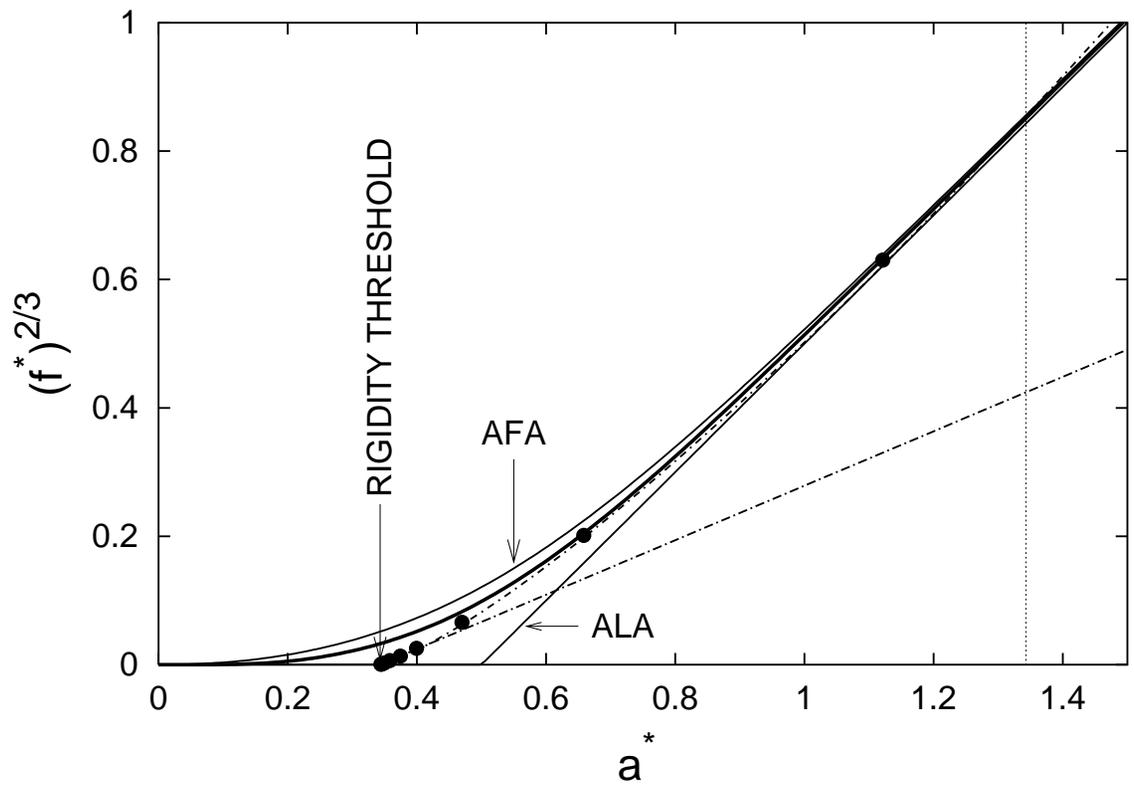
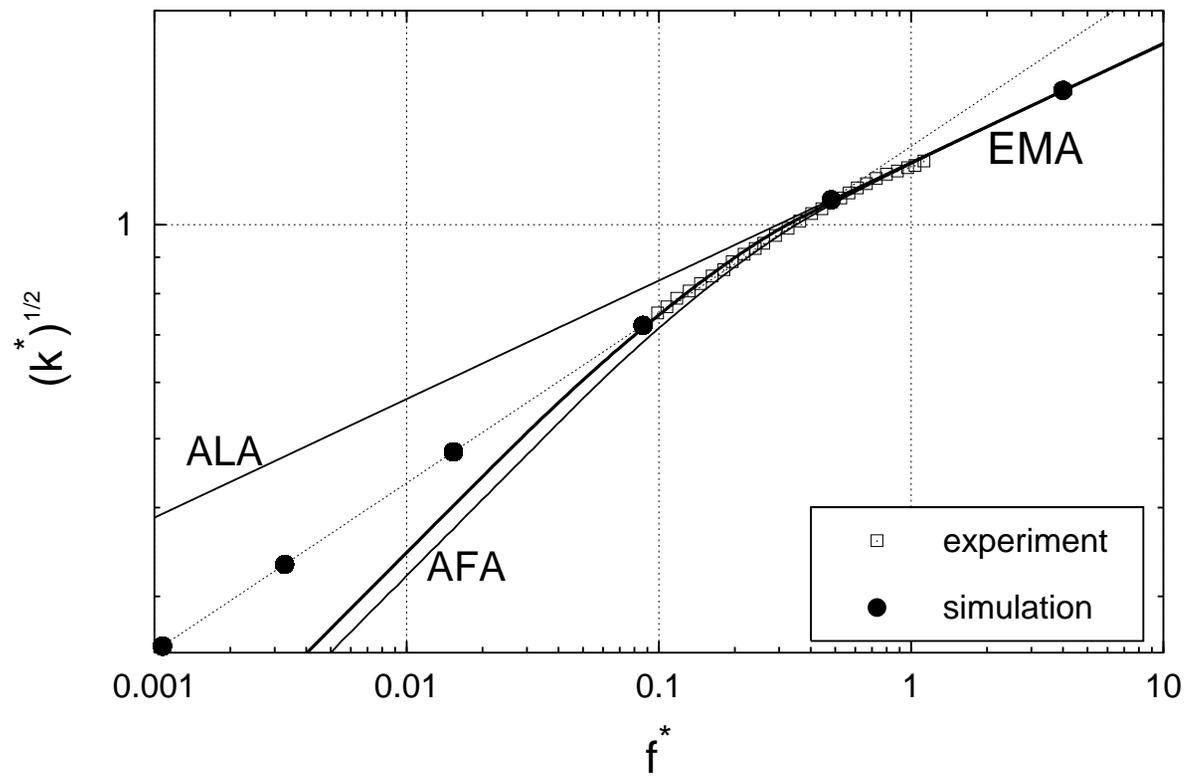
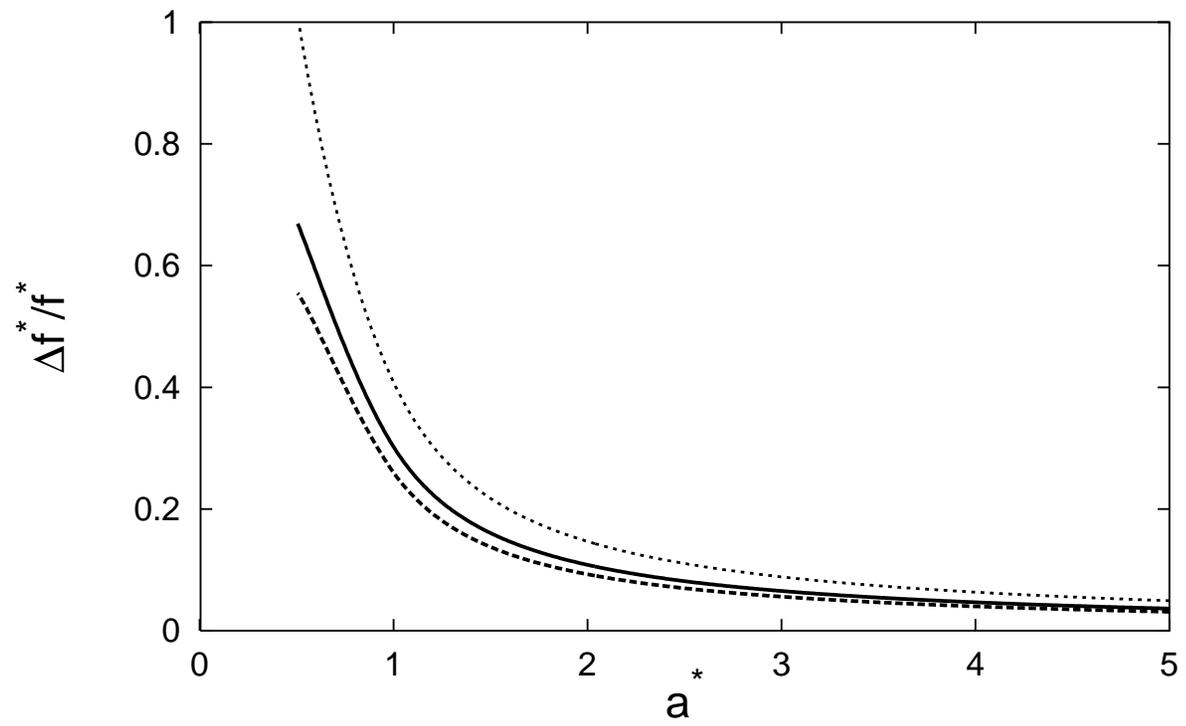
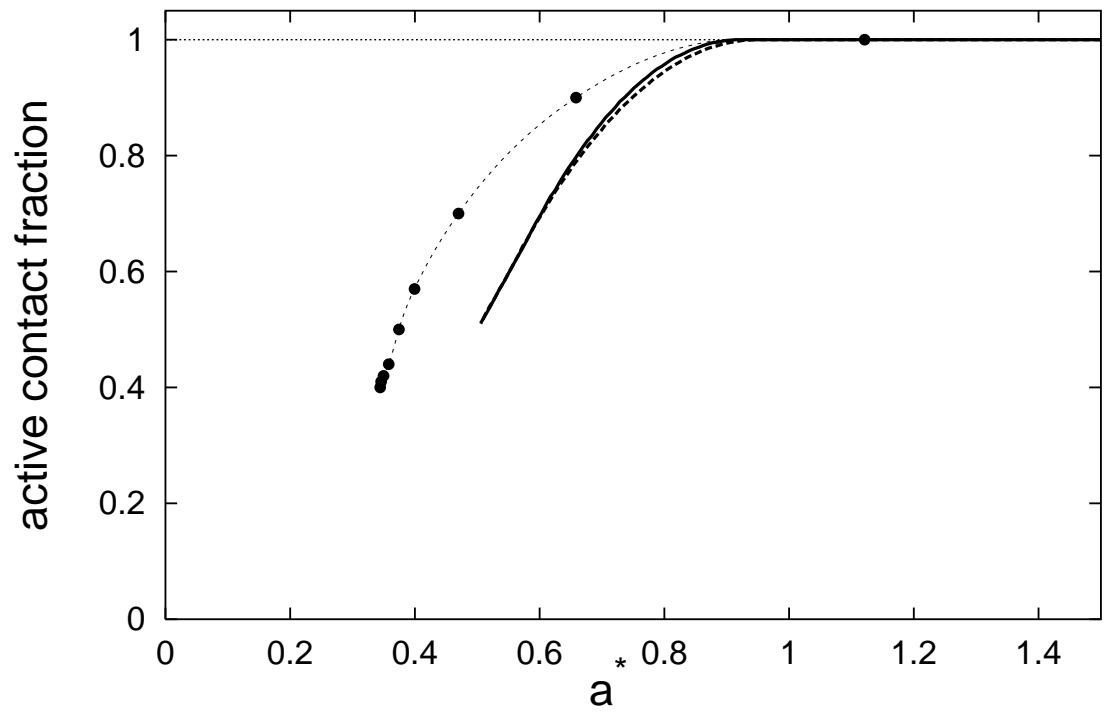
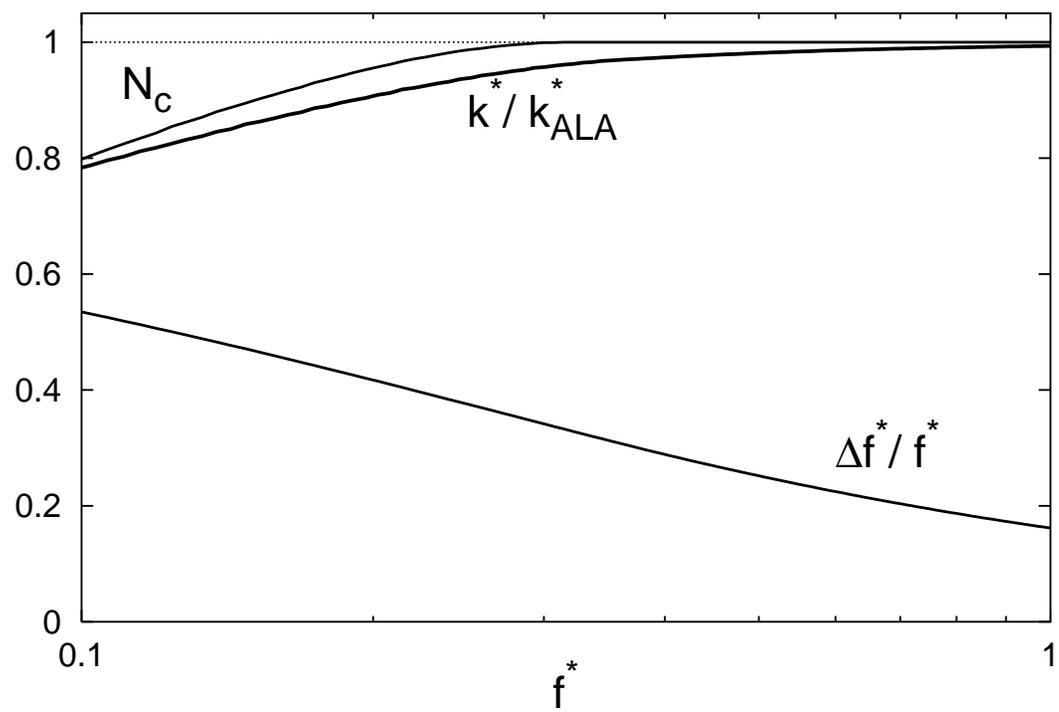
\end{document}